\documentclass[usenatbib]{mn2e}

\usepackage{epsfig}

\title[Abundance Spread in To 2?]{Tombaugh 2: The First Open Cluster with a Significant Abundance Spread or 
Embedded in a Cold Stellar Stream?}
\author[Frinchaboy, P.M., et al.]{P. M. Frinchaboy$^{1}$\thanks{E-mail: pmf@astro.wisc.edu (PMF);
anna.marino@unipd.it (AFM); sandro.villanova@unipd.it (SV); gcarraro@eso.org (GC);   
srm4n@virginia.edu (SRM); dgeisler@astro-udec.cl
(DG)}
\thanks{Any opinions,
findings, and conclusions or recommendations
expressed in this material are those of the
author(s) and do not necessarily reflect the views of the National Science
Foundation.}
\thanks{Based on observations collected at the
European Southern Observatory, Chile; Proposal 076.B-0263},
A. F. Marino$^{2}$, S. Villanova$^{2,3}$, G. Carraro$^{2,4}$, \newauthor S. R. Majewski$^{5}$, and D. Geisler$^{3}$\\
$^{1}$National Science Foundation Astronomy \& Astrophysics
Postdoctoral Fellow, Univeristy of Wisconsin--Madison, \\
\indent Department of Astronomy, 4506 Sterling Hall,
475 N.\ Charter Street, Madison, WI 53706, USA\\
$^{2}$Dipartimento di Astronomia, Universit\`a di Padova, Vicolo Osservatorio 5, I-35122 Padua, Italy\\
$^{3}$Universidad de Concepci\'on, Departamento de Fisica,
Casilla 160-C, Concepci\'on, Chile\\
$^{4}$ESO, Alonso de Cordova 3107, Vitacura, Santiago, Chile\\
$^{5}$Department of Astronomy, University of Virginia,
P.O. Box 400325, Charlottesville, VA 22904-4325, USA
}
\begin{document}

\date{Accepted 2008 January xx. Received 2008 January xx; in original form 2008 January}
\pagerange{\pageref{firstpage}--\pageref{lastpage}} \pubyear{2007}

\maketitle
\label{firstpage}

\begin{abstract}
We present new high resolution spectroscopy from which we derive 
abundances and radial velocities for stars in the field of
the open cluster Tombaugh 2, which has been suggested to be one of a group of
clusters previously identified with the 
Galactic Anticenter Stellar Structure (also known as the Monoceros stream). 
Using VLT/FLAMES with the UVES and GIRAFFE spectrographs,
we find a radial velocity (RV) of $\langle V_{r}\rangle = 121 \pm 0.4$
km s$^{-1}$ using eighteen Tombaugh 2 cluster stars; this is in agreement with
previous studies, but at higher precision.
We also make the first measurement of Tombaugh 2's 
velocity dispersion, which is $\sigma_{int} = 1.8 \pm 0.3$ km s$^{-1}$.
Our abundance analysis of RV-selected members finds that Tombaugh 2 is
more metal-rich than previous studies have found; 
moreover, unlike the previous work, our larger sample
also reveals that stars with the velocity of the cluster 
show a relatively large spread in chemical properties 
(e.g., $\Delta$[Fe/H] $> 0.2$).
This is
the first time a possible abundance spread has been observed in an {\it open}
cluster, though this is one of several possible explanations for our
observations.
While there is an apparent trend of [$\alpha$/Fe] with [Fe/H], 
the distribution of abundances of these ``RV cluster
members'' also may hint at a possible division into two primary groups 
with different mean chemical characteristics ---
namely ($\langle$[Fe/H]$\rangle$, $\langle$[Ti/Fe]$\rangle$) $\sim$
($-$0.06, $+$0.02) and ($-$0.28, $+$0.36).
Isochrone fitting to the colour-magnitude distribution of apparent Tombaugh 2 members yields an
age of 2.0 Gyr, $E(B-V) = 0.3$, and $(m-M)_0 = 14.5$ or $d = 7.9$ kpc for both populations --- 
parameters that are within the range of previous findings.
Based on position and kinematics Tombaugh 2 is a likely member of the GASS/Monoceros stream, 
which makes Tombaugh 2 the second star cluster within the originally
proposed GASS/Monoceros family after NGC2808 to show some evidence
for internal population dispersions.  However, we explore other
possible explanations for the observed spread in abundances and
two possible sub-populations, with the most likely explanation being
that the metal-poor ([Fe/H] = $-$0.28), more centrally-concentrated 
population being the true Tombaugh 2 clusters stars and the 
metal-rich ([Fe/H] = $-$0.06) population being an overlapping,
and kinematically associated, but ``cold''  ($\sigma_V < 2$ km s$^{-1}$) stellar stream at $R_{gc} \ge 15$ kpc.
 \end{abstract}


\section{Introduction}

Tombaugh 2 (To2; at Galactic coordinates [$l,b$] = [232.8,$-$6.9]$^{\circ}$)
is an old, distant open cluster in the outer Galactic disc.
We originally targeted this cluster 
as part of an ongoing program \citep[e.g.,][]{carraro07}
to explore in depth a set of star clusters proposed by \citet{pmf04} ---
on the basis of their position and kinematics --- to be possibly
associated with the Monoceros stream (Mon; Newberg et al. 2002, Ibata et al. 2003, 
Yanny et al. 2003), 
also known as the Galactic anticenter stellar structure (GASS; Rocha-Pinto et al 2003, 
Crane et al. 2003).  
GASS/Mon was discovered as an overdensity of stars near the Galactic plane
that seems to wrap around the outer parts of the Galactic disc and is frequently
explained as tidal debris from the disruption of a dwarf galaxy on a low inclination orbit
(e.g., Crane et al. 2003, Yanny et al. 2003,  Penarrubia et al. 2005).
To2 has also been associated with
the reputed dwarf galaxy in Canis Major (CMa; Bellazzini et al. 2004), which
is suggested to be a progenitor of the Mon/GASS structure (Martin et al. 2004a).  
However, Rocha-Pinto et al. (2006)
argue that a more likely progenitor of Mon/GASS may lie somewhere 
in the region of the sky covered by the former constellation Argo Navis, and that
the overdensity reported near CMa is correlated to a window in the dust extinction there;
in this case, To2 would be farther displaced from the putative Mon/GASS progenitor.
In any of these circumstances To2 is interesting as a potential member of a star cluster 
system accreted from another galaxy by the Milky Way.  

Subsequently, our spectroscopic analysis of this cluster reveals
To 2 to be interesting in its own right as the first known open cluster exhibiting possible evidence
for multiple populations or an internal population spread, evidence we present and explore here
No well studied open cluster has shown a spread in abundances of
major elements \citep[e.g.,][study of M67]{Randich06}. 
The possible association of To2 to Mon is a characteristic shared with another
unusual star cluster showing multiple populations, the 
globular cluster NGC 2808 (Crane et al. 2003, Martin et al. 2004a).\footnote{It should be noted that 
Casetti-Dinescu et al. (2007) have excluded NGC 2808 from being a part of the Canis Major/Mon structures based 
on the model of Pe\~narrubia et al. (2005).  However if Monoceros/GASS and Canis Major are 
not related or if NGC 2808 is like $\omega$ Cen 
(as suggested by Casetti-Dinescu et al. 2007) and has its own tidal stream, the derived NGC 2808 
orbit remains consistent with the tilted spatial configuration found in outer disc star clusters 
(Frinchaboy et al. 2004).}  
But whereas a few other globular clusters with multiple populations are now known
(see \S6.2), to date no multi-population open clusters have been reported.  Nevertheless, 
having two unusual, multi-population clusters potentially associated with the GASS/Mon/CMA/Argo
structures is intriguing and may point to a mechanism for their creation ... if we can first establish
that association {\it and} determine what GASS/Mon/CMA/Argo is.

The nature and/or reality of the proposed Mon ``tidal debris stream'' and the CMa 
overdensity have been called 
into question and are currently a matter of great debate. 
Ibata et al. (2003) originally proposed that Mon could be related to warps in the outer disc, while
Momany et al. (2004, 2006) have   
argued that much of the observed stellar 
overdensity associated with Mon --- and particularly all of that associated with 
CMa (at $l \sim 240^{\circ}$) ---
is due to the warping and flaring of the Galactic disc, and that no ``extra-Galactic'' component is 
needed to account for the apparent overdensities in the third quadrant.  This conclusion
has been contested by Martin et al. (2004b) on the basis of radial velocities (but cf. 
Momany et al. 2006), while the discovery of  ``blue plume'' stars in this part of the sky
has been used to argue further for the presence of a dwarf galaxy nucleus in CMa
(Bellazzini et al. 2004, Martinez-Delgado et al. 2005, Dinescu et al. 2005, Butler et al. 2007).
However, these young stars have also been more prosaically attributed to the presence of spiral arm
structure (Carraro et al. 2005, Moitinho et al. 2006).  And Grillmair (2006) has quite clearly identified 
tidal streams across the Galactic anticenter region
with relatively low inclination (35$^{\circ}$) and at a similar distance and main sequence turnoff colour
as Mon, though he concludes on the basis of the best fit orbits that these streams have nothing to do with Mon
or CMa.  

What is clear from all of the debate, and a commonly expressed sentiment by all sides,
is that further study, particularly detailed spectroscopic analysis,
is needed before the complicated
nature of the outer disc, its warp, flare, spiral arm structure and possibly co-located tidal debris
and halo substructure, can be confidently disentangled.
Thus, a number of studies \citep[e.g.,][]{yong04,yong2,pmf05} are being 
conducted to explore further the kinematical and chemical properties 
of Monoceros and the outer Galactic disc.  Our own venture in this regard
starts by focusing on high resolution spectroscopic investigation of old, open star clusters 
of the outer disc, including those hypothesized 
to be parts of the debated overdensities.  In \citet{carraro07} we present
high resolution spectroscopy of the first five outer disc open
clusters observed in our survey (Ruprecht~4, Ruprecht~7, Berkeley~25, Berkeley~73 and
Berkeley~75) and derive detailed abundance analyses and kinematics for these systems.
Here, we discuss separately the case of the open cluster To2, which has surfaced as 
a potentially unusual system among the clusters of the outer disc.
In this paper we obtain new estimates of the radial velocities (\S 3) and 
abundances ([Fe/H] and [Ti{\hspace{0.1cm}\scriptsize\rm I}/Fe]; \S 4) of stars in the To2 field.  We discuss 
the unusual abundance results for To2 in \S5, and address the possible origins of the observed
mixed chemistry,
its implications for To2 and the outer disc, and the issue of the 
possible connection of To2 to the Monoceros/GASS structure in \S6.

\begin{figure*}
\includegraphics[width=170mm]{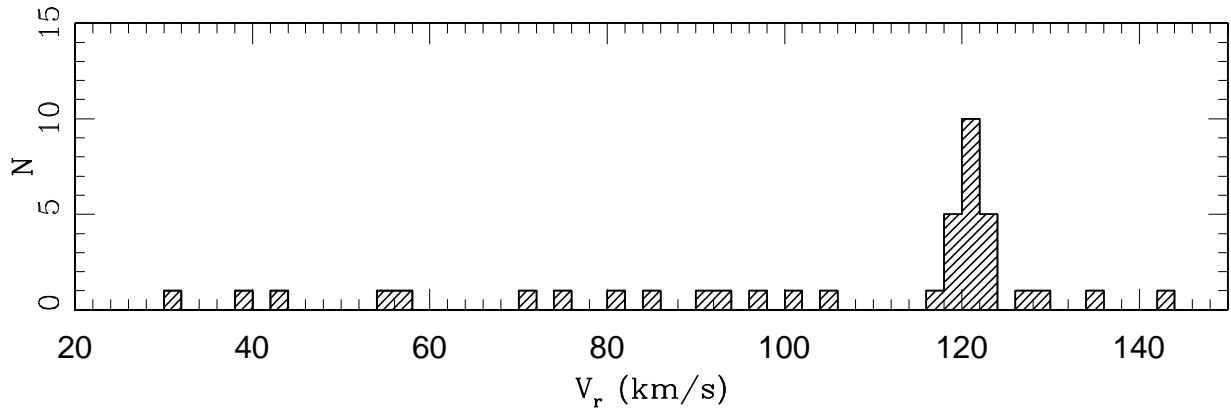}
\caption{Radial velocity distribution of all stars observed with the VLT.  
The cluster is clearly visible as the
peak near $\sim 121$ km s$^{-1}$.\label{rvhist}
}
\end{figure*}

\section{Observations and Data Reduction}

Our spectroscopic data of To2 come from the ESO-VLT (proposal 076.B-0263),
and were collected 
with the FLAMES@VLT+GIRAFFE \citep{flames} and FLAMES@VLT\-+UVES \citep{uves}
spectrographs on the nights of UT 1-Dec-2005 and 5-Dec-2005.
The sky
was clear, and the typical seeing was 1.0 arcsec.
FLAMES  allows the use of the GIRAFFE spectrograph to obtain RVs for up to 132 stars
at one time,
while simultaneously using the UVES
multi-fiber mode, which allows observing up to eight additional sources.
For UVES, we used the 580nm set-up ($R=40,000$ in the 4750--6800 \AA~range).
The eight fibers were placed on the brighter probable members of the 
cluster (selected according to their position in the 
colour-magnitude diagram; CMD), while FLAMES+GIRAFFE was used to target 35 additional probable members,
with the remaining FLAMES+GIRAFFE fibers sampling the sky.  
In Table \ref{rvs} we report properties of the observed stars: the first and
second column give their ID, the third and fourth the coordinates, the
fifth and sixth the $V$ magnitude and $V-I$ colour, the seventh,
eighth, and ninth columns the heliocentric radial velocity with the error and
the membership classification (see below),
while the last three columns give the adopted atmospheric parameters
 used for abundance measurements for the member stars (see \S 3.2).

The data were reduced by ESO personnel using the FLAMES-UVES reduction pipeline
(see http://www.eso.org/projects/dfs/dfs-shared/web/vlt/vlt-instrument-pipelines.html
for documentation on the FLAMES-UVES pipeline and software)
which corrects the spectra for the detector bias and flat-field. Then
a wavelength calibration based on Th-Ar calibration-lamp spectra
was applied. Finally, each spectrum was flux-calibrated by applying
the response-curve of the instrument, and the echelle orders were combined
to obtain a single mono-dimensional spectrum.
The resulting UVES spectra have a dispersion of 0.1 \AA~pixel$^{-1}$ and a typical
$S/N\sim$ 15-20.

\begin{table*}
 \centering
 \begin{minipage}{140mm}
  \caption{Radial velocities, coordinates, photometry,
  cluster membership, and atmospheric parameters for stars with sufficient 
  $S/N$ to derive abundances from the GIRAFFE and UVES stellar spectra.}
  \label{rvs}
  \begin{tabular}{rlllccrccccc}
  \hline
ID & ID2 & RA(2000.0) & DEC(2000.0) & $V$ & $V$-$I$ & RV$_{helio}$ & $\epsilon_{RV}$ & Member? & T$_{\it eff}$ & log(g) & v$_{\it t}$ \\
 \hline
\multicolumn{12}{c}{\underline{GIRAFFE spectra}}\\
63  &  20 &  07:03:06.837 &  $-$20:48:28.34 & 15.078 &  1.505 &  120.96 &  0.16 &  Y  & 4800 & 2.65 & 2.00\\
76  &     &  07:03:23.038 &  $-$20:53:39.11 & 15.327 &  1.479 &   84.90 &  0.18 &  N  & - & - & - \\
80  &     &  07:03:24.743 &  $-$20:48:13.52 & 15.340 &  1.543 &   54.19 &  0.26 &  N  & - & - & - \\
89  &     &  07:03:15.392 &  $-$20:52:26.04 & 15.455 &  1.446 &   90.60 &  0.16 &  N  & - & - & - \\
96  &     &  07:03:20.364 &  $-$20:45:55.79 & 15.541 &  1.453 &  127.73 &  0.14 &  N  & - & - & - \\
97  &     &  07:03:19.745 &  $-$20:51:52.63 & 15.553 &  1.656 &  105.83 &  0.32 &  N  & - & - & - \\
98  &  28 &  07:03:09.216 &  $-$20:51:29.22 & 15.556 &  1.573 &  121.71 &  0.18 &  Y  & 4930 & 2.85 & 2.00 \\
102 &  31 &  07:03:06.548 &  $-$20:49:36.95 & 15.634 &  1.295 &  121.99 &  0.18 &  Y  & 5200 & 2.35 & 1.70 \\
106 &     &  07:03:21.757 &  $-$20:54:08.42 & 15.683 &  1.227 &   43.39 &  0.36 &  N  & - & - & - \\
107 &     &  07:02:53.866 &  $-$20:46:39.40 & 15.691 &  1.294 &   93.21 &  0.16 &  N  & - & - & - \\
109 &     &  07:03:13.691 &  $-$20:52:58.65 & 15.705 &  1.408 &   74.43 &  0.22 &  N  & - & - & - \\
110 &  36 &  07:03:06.097 &  $-$20:50:31.00 & 15.721 &  1.530 &  122.07 &  0.28 &  Y  & - & - & - \\
115 &     &  07:03:12.518 &  $-$20:49:27.04 & 15.759 &  1.225 &   38.97 &  0.50 &  N  & - & - & -\\
124 &  38 &  07:03:07.271 &  $-$20:50:01.05 & 15.897 &  1.505 &  121.79 &  0.18 &  Y  & 5000 & 3.20 & 1.80 \\
126 &  39 &  07:02:55.173 &  $-$20:49:21.06 & 15.915 &  1.361 &  120.58 &  0.18 &  Y  & - & - & - \\
127 &     &  07:02:55.451 &  $-$20:51:15.67 & 15.919 &  1.468 &  121.10 &  0.18 &  Y  & 4850 & 3.50 & 1.30 \\
135 &     &  07:02:53.903 &  $-$20:50:09.95 & 15.947 &  1.443 &  119.18 &  0.18 &  Y  & 4900 & 3.17 & 2.00 \\
140 &  42 &  07:02:59.716 &  $-$20:49:33.60 & 15.981 &  1.454 &  121.16 &  0.18 &  Y  & 5250 & 3.25 & 1.40 \\
141 &     &  07:02:51.099 &  $-$20:47:15.37 & 15.989 &  1.285 &  104.04 &  0.18 &  N  & - & - & - \\
142 &     &  07:03:27.918 &  $-$20:52:19.99 & 15.992 &  1.294 &  101.07 &  0.16 &  N  & - & - & - \\
177 &  61 &  07:03:07.185 &  $-$20:50:20.63 & 16.206 &  1.395 &  118.74 &  0.16 &  Y  & 5070 & 3.55 & 2.00 \\
179 &  63 &  07:03:13.226 &  $-$20:49:42.27 & 16.215 &  1.364 &  123.73 &  0.18 &  Y  & 5000 & 3.50 & 1.75 \\
182 &  65 &  07:03:02.625 &  $-$20:48:23.84 & 16.256 &  1.355 &  120.87 &  0.16 &  Y  & 5150 & 3.57 & 1.70\\
190 &     &  07:02:52.311 &  $-$20:44:38.53 & 16.290 &  1.222 &   80.56 &  0.16 &  N  & - & - & -\\
191 &  67 &  07:03:05.065 &  $-$20:48:57.77 & 16.296 &  1.389 &  121.20 &  0.18 &  Y  & 4980 & 2.55 & 2.05 \\
192 &     &  07:02:53.283 &  $-$20:48:01.52 & 16.301 &  1.391 &   96.34 &  0.20 &  N  & - & - & - \\
196 &  70 &  07:03:06.455 &  $-$20:49:17.02 & 16.322 &  1.466 &  123.57 &  0.18 &  Y  & 4950 & 3.15  & 2.10 \\
199 &  71 &  07:03:04.003 &  $-$20:49:08.06 & 16.341 &  1.407 &  121.53 &  0.16 &  Y  & 5050 & 2.85  & 1.20 \\
201 &     &  07:02:57.371 &  $-$20:52:58.36 & 16.351 &  1.336 &   57.57 &  0.16 &  N  & - & - & - \\
207 &     &  07:03:19.387 &  $-$20:48:39.75 & 16.387 &  1.475 &   31.98 &  0.92 &  N  & - & - & - \\
215 &  74 &  07:03:11.214 &  $-$20:49:33.24 & 16.454 &  1.409 &  128.76 &  0.20 &  N  & - & - & - \\
217 &  75 &  07:03:09.103 &  $-$20:49:25.97 & 16.459 &  1.240 &  151.93 &  0.56 &  N  & - & - & - \\
219 &     &  07:02:48.848 &  $-$20:44:11.12 & 16.468 &  1.334 &  134.07 &  0.18 &  N  & - & - & - \\
223 &     &  07:02:48.365 &  $-$20:47:37.97 & 16.523 &  1.238 &   70.62 &  0.42 &  N  & - & - & - \\
231 &     &  07:03:28.596 &  $-$20:46:26.52 & 16.580 &  1.459 &  122.11 &  0.14 &  Y  & 4980 & 3.20 & 2.00 \\
\multicolumn{12}{c}{\underline{UVES spectra}}\\
146 &  44 &  07:03:09.688 &  $-$20:45:49.21 & 16.018 &  1.367 &  119.55 &  0.02 &  Y  & - & - & - \\
158 &  50 &  07:03:07.759 &  $-$20:46:17.44 & 16.099 &  1.341 &  118.78 &  0.03 &  Y  & 5330 & 2.35 & 2.01 \\
162 &  52 &  07:03:01.566 &  $-$20:47:59.76 & 16.131 &  1.318 &  118.01 &  0.49 &  Y  & 5290 & 2.72 & 1.90 \\
164 &     &  07:03:20.480 &  $-$20:46:47.78 & 16.160 &  1.436 &  122.62 &  0.14 &  Y  & 5270 & 3.56 & 1.39 \\
165 &  55 &  07:03:03.496 &  $-$20:48:48.42 & 16.161 &  1.411 &  117.23 &  0.03 &  Y  & 5190 & 2.51 & 1.71 \\
\hline
\end{tabular}
\end{minipage}
\end{table*}

The GIRAFFE spectra were obtained simultaneously using the HR09B set-up ($R=21,000$ in the 5138--5350 \AA~range)
with 35 stars targeted.  These data 
were reduced similarly to the UVES spectra, 
are at a dispersion of 0.25\AA~pixel$^{-1}$, 
and have $S/N\sim$ 60--70.

The stars targeted with GIRAFFE and UVES
were observed twice, so that two independent RV measurements were derived
for each star
from the collected spectra.
Radial velocities were measured using the IRAF utility {\tt fxcor};
this routine cross-correlates the observed spectrum with a template
having known radial velocity.
As a template, we used a synthetic spectrum calculated
for a typical solar metallicity giant star [$T_{\rm{eff}}\sim 5000$ K,
$log(g)=3.0$, $v_t=1.3$ km s$^{-1}$].
Then each {\tt fxcor} measured RV was converted to a heliocentric velocity
using the IRAF routine {\tt rvcorr}.
All of the measured RVs are presented in Table \ref{rvs}
with their errors; RVs for all observed stars are the weighted mean from the two
measurements for each star, and these provide estimates of the errors by taken from {\tt fxcor}.

\section{Membership and Cluster Kinematics}\label{model}

The derived radial velocities and metallicities,
along with the available photometry of To2 stars \citep*{kubiak92,phelps94}, are used to 
revise the fundamental parameters of the cluster. Cluster membership
for the observed stars was determined on the basis of their radial velocity (Table \ref{rvs}).
The heliocentric velocities for all the observed stars ($V_{r}$) are reported in 
Table \ref{rvs}, which provides IDs from \citet[][ col. 1]{phelps94} and \citet[][ col. 2]{kubiak92},
stellar coordinates, magnitudes and colours from \citet{phelps94}.
As shown in Figure \ref{rvhist}, we can easily identify cluster stars from their RV.  
Stars with radial velocities of 121 $\pm$ 4 km s$^{-1}$ are considered to be cluster members 
on the basis that: (1) examination of the RV distribution in Figure \ref{rvhist} clearly shows a peak 
associated with the cluster at 121 km s$^{-1}$, (2) open cluster velocity dispersions are well-known
to be very small, typically less than 0.5--1 km s$^{-1}$ 
\citep[e.g.,][]{liu91,girard89,gim98,deB01,meibom02,hole03}.
Thus, the $\pm$ 4 km s$^{-1}$ acceptance window for members is more than generous
for including suspected RV members, but still small enough to exclude 
most (but possibly not all) expected Galactic contamination.   From the distribution
of RVs shown in Figure 1, one might expect field star contamination across an 8 km s$^{-1}$ 
window at the level of no more than a few stars.  
Moreover, from the Besancon model \citep{bec_model} we expect the mean velocity of Milky Way stars 
along the To2 line of sight to be about 57 km s$^{-1}$, with a dispersion of 38 km s$^{-1}$;
which suggests, given our sampling, a contamination rate of 0.02 stars ($\sim2$\%) with RVs of 117--125 km s$^{-1}$
within the observed field of view.

\begin{figure*}
\includegraphics[width=170mm]{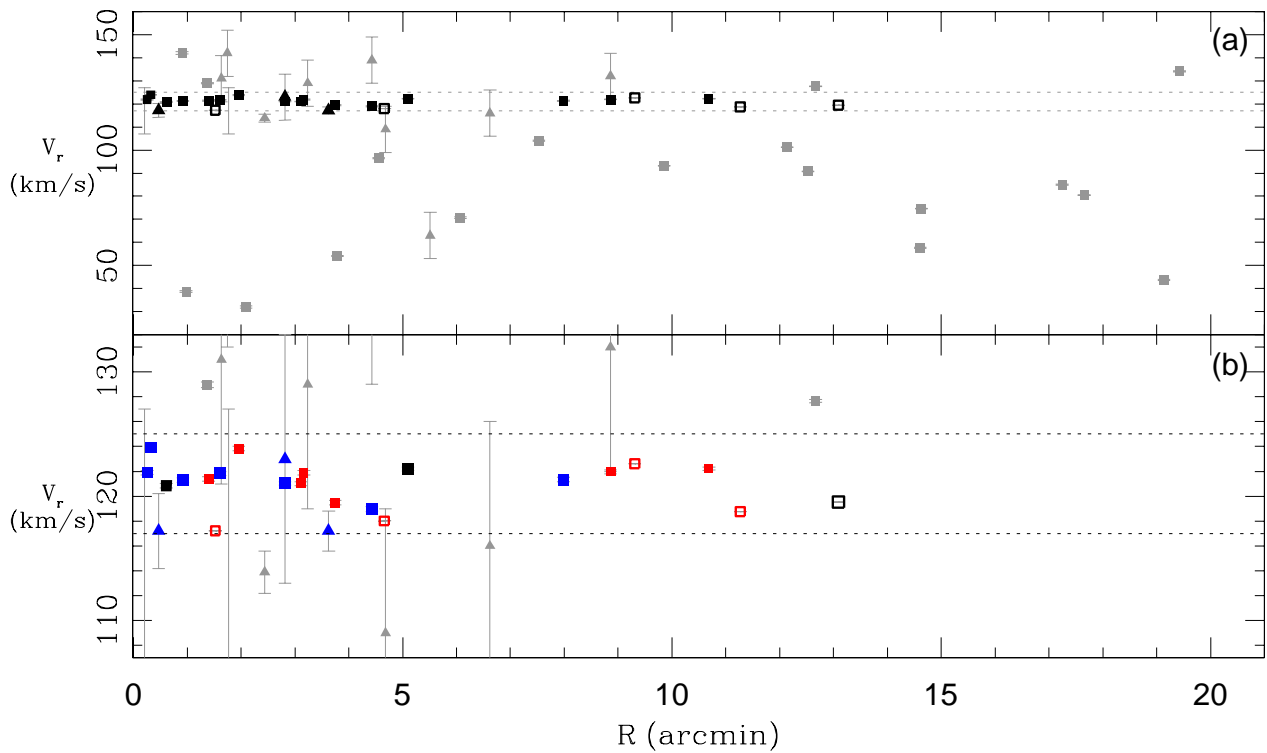}
\caption{(a) The distribution of all new and previously measured radial velocities (with error bars) for 
	stars in the field of Tombaugh 2 as a function of their projected distance from the cluster centre.  
         Squares denote stars observed
         in this study, with filled squares denoting those observed with GIRAFFE and open squares
         	representing those with UVES spectra.
         Stars selected to be RV members are shown by black symbols.
        Triangles denote RVs 
         from \citet{brown96}
         and \citet{friel02}, with darker triangles denoting members from these studies
         by our RV definition.
         (b) Same as (a) 
         but with metallicity information for member stars shown by colour: red points are
         ``metal-rich'' ([Fe/H] $> -$0.15) and blue points are ``metal-poor''
         ([Fe/H] $\le -$0.15) --- see the discussion in \S \ref{revise}.
         \label{rv1}
}
\end{figure*}

Analyzing the RV membership sample (shown within the dotted lines in Figure \ref{rv1}) 
using techniques from \citet{pm93},
we derive a systemic RV for To2 of $\langle V_{r}\rangle$ = 121.04 $\pm$ 0.43 km s$^{-1}$
and a total velocity dispersion 
of $\sigma_{int}$ = 1.81 $\pm$ 0.30 km s$^{-1}$.  
Of course, this dispersion may be somewhat inflated by any field star contamination and/or motions
of unresolved binary stars.
Cluster membership and the To2 velocity dispersion is addressed further in \S \ref{revise}.

\section{Abundance analysis}

\subsection{Atomic parameters and equivalent widths}

The analysis of chemical abundances was carried out with the latest
version (2005) of the program MOOG \citep{sneden73} and using model
atmospheres by \citet{kurucz79}. MOOG performs its analysis in local thermodynamic
equilibrium (LTE).
We derived equivalent widths of unblended spectral lines
by using a semi-automatic SuperMongo program 
written by one of
the authors. Repeated measurements show a
typical error of about 5-10m\AA~for the weakest lines
because of the relatively low $S/N$ of the spectra
(UVES: $\sim20$, GIRAFFE: 60--70 ) compared to those typically used in chemical abundance studies.
The line list was taken from \citet{gratton03} for UVES spectra,
and from the VALD\footnote{Available at http://www.astro.uu.se/$\sim$vald/} 
(Vienna Atomic Line Database; Kupka et al. 1999) database for GIRAFFE spectra.
The log($gf$) parameters of these lines were redetermined by a solar-inverse
analysis measuring the equivalent widths from the NOAO solar spectrum and adopting the
standard solar parameters [$T_{eff}$ = 5777 K, log($g$) = 4.44, and $v_t$ = 0.8 km s$^{-1}$].
The solar abundances found by this analysis are reported in Table \ref{solar}.  
For each element we give also the number of measured
spectral lines. In one case (Ni{\hspace{0.1cm}\scriptsize\rm I}) the solar abundances for
UVES and GIRAFFE spectra differ by more than 0.2 dex. The reason is
that we decided to adjust the log($gf$) values only to
remove the scatter affecting each line list
and not to register the two line lists onto a common scale;
as a result there remains a possible systematic offset for
Ni{\hspace{0.1cm}\scriptsize\rm I}.

\begin{table}
 \centering
 \begin{minipage}{80mm}
  \caption{Adopted solar abundances for UVES and GIRAFFE spectra.}
\label{solar}
  \begin{tabular}{llrlr}
  \hline
Element & UVES &\# lines   & GIRAFFE& \# lines \\
 \hline
Fe{\hspace{0.1cm}\scriptsize\rm I}  & 7.48     &  50  & 7.51  & 40  \\
Fe{\hspace{0.1cm}\scriptsize\rm II} & 7.51     &   4  & 7.52  &  6  \\
Na{\hspace{0.1cm}\scriptsize\rm I}  & 6.31(LTE)&   2  &       &     \\
Mg{\hspace{0.1cm}\scriptsize\rm I}  & 7.53     &   1  &       &     \\
Si{\hspace{0.1cm}\scriptsize\rm I}  & 7.61     &   1  &       &     \\
Ca{\hspace{0.1cm}\scriptsize\rm I}  & 6.37     &   2  & 6.55  &  2  \\
Ti{\hspace{0.1cm}\scriptsize\rm I}  & 4.93     &   9  & 5.08  &  7  \\
Ti{\hspace{0.1cm}\scriptsize\rm II} & 4.96     &  11  & 5.08  &  3  \\
Cr{\hspace{0.1cm}\scriptsize\rm I}  & 5.65     &   5  & 5.66  &  7  \\
Cr{\hspace{0.1cm}\scriptsize\rm II} & 5.72     &   2  &       &     \\
Ni{\hspace{0.1cm}\scriptsize\rm I}  & 6.26     &  11  & 6.51  &  4  \\
Ba{\hspace{0.1cm}\scriptsize\rm II} & 2.45     &   2  &       &     \\
\hline
\end{tabular}
\end{minipage}
\end{table}

  \begin{table}
   \centering
   \begin{minipage}{80mm}
    \caption{Sensitivity of derived abundances to the atmospheric
  parameters.\label{errors}}
    \begin{tabular}{lccc}
    \hline
          & $\Delta(T)$ &$\Delta(\log(g))$ & $\Delta(v_{t})$\\
  Element & (per +100 K) & (per +0.2) & (per +0.2 km s$^{-1}$)\\
   \hline
  Fe{\hspace{0.1cm}\scriptsize\rm I}  & $+$0.07     & $+$0.02 & $-$0.07\\
  Fe{\hspace{0.1cm}\scriptsize\rm II} & $-$0.07     & $+$0.10 & $-$0.06\\
  Na{\hspace{0.1cm}\scriptsize\rm I}  & $+$0.08     & $-$0.04 & $+$0.04\\
  Mg{\hspace{0.1cm}\scriptsize\rm I}  & $-$0.08     & $-$0.04 & $+$0.06\\
  Si{\hspace{0.1cm}\scriptsize\rm I}  & $-$0.02     & $+$0.02 & $+$0.05\\
  Ca{\hspace{0.1cm}\scriptsize\rm I}  & $+$0.09     & $-$0.04 & $-$0.09\\
  Ti{\hspace{0.1cm}\scriptsize\rm I}  & $+$0.15     & $-$0.02 & $-$0.11\\
  Ti{\hspace{0.1cm}\scriptsize\rm II} & $-$0.01     & $+$0.10 & $+$0.12\\
  Cr{\hspace{0.1cm}\scriptsize\rm I}  & $+$0.13     & $-$0.05 & $-$0.10\\
  Cr{\hspace{0.1cm}\scriptsize\rm II} & $-$0.04     & $+$0.09 & $-$0.10\\
  Ni{\hspace{0.1cm}\scriptsize\rm I}  & $-$0.08     & $+$0.00 & $+$0.11\\
  Ba{\hspace{0.1cm}\scriptsize\rm II} & $+$0.03     & $+$0.05 & $-$0.15\\
  \hline
  \end{tabular}
  \end{minipage}
  \end{table}

\subsection{Atmospheric parameters}\label{atmparam}

Initial estimates of the atmospheric parameter $T_{\rm{eff}}$ were
obtained from photometric observations using the relations from \citet{alonso99}.
We first adopted $E(B-V)$ = 0.35 \citep{brown96} to correct colours for the interstellar extinction.
We then adjusted the effective temperature to minimize the slope of
the abundances obtained from Fe I lines with respect to the excitation
potential in the curve of growth analysis.
Initial guesses for the gravity, $\log{g}$ were derived from the canonical
formula:
\begin{equation}
\log\left(\frac{g}{g_{\odot}}\right) =
\log\left(\frac{M}{M_{\odot}}\right) + 4
\log\left(\frac{T_{\rm{eff}}}{T_{\odot}}\right)
- \log\left(\frac{L}{L_{\odot}}\right)
\end{equation}
In this equation the mass $M/M_{\odot}$ was derived from the
comparison between the position of the star in the colour-magnitude diagram
and the Padova isochrones \citep{girardi00}.
The luminosity $L/L_{\odot}$ was derived from the absolute magnitude
$M_V$,  assuming a distance modulus of $(m-M)_V=15.3$ \citep{kubiak92}. The bolometric correction
(BC) was derived from the BC-$T_{\it eff}$ relation from \citet{alonso99}.
The input $\log{g}$ values were then adjusted in order to satisfy the
ionization equilibrium of Fe I and Fe II during the abundance
analysis.
Finally, the microturbulence velocity is given by the relation
\citep*{hou00}:
\begin{equation}
v_{\rm t} = 2.22 - 0.322 \log g
\end{equation}
We then adjusted the microturbulence velocity by minimizing the slope
of the abundances obtained from Fe I lines with respect to the
equivalent width in the curve of growth analysis.  
From the derived spectroscopic temperatures, we are able to obtain
the intrinsic $(B-V)$ colours for our stars using \citet{alonso99}.
We find that for these stars that $E(B-V) \sim 0.30 \pm 0.04$,
in agreement with 
the reddening estimate for To2 by \citet{brown96}.
The final derived stellar parameters are listed in Table
\ref{rvs}. The uncertainties in the spectroscopically determined
stellar parameters are of the order of $\pm$100 K in $T_{eff}$,
$\pm$0.2 in $log(g)$, and $\pm$0.2 km s$^{-1}$ in microturbolent
velocity for red giant stars \citep[see][]{friel05}.  
Table \ref{errors} lists the impact of uncertainties in the
atmospheric parameters on the derived abundances for the elements
considered in our analysis.
Variations in parameters of the model atmospheres were obtained by
changing each of the parameters one at a time for one of the analyzed
stars (\#179), assumed to be representative of all the stars considered in this
paper.

\subsection{Stellar abundances}

The derived abundances from the UVES spectra are presented in Table \ref{Uabund1}. 
In addition to the UVES spectra, we also derive some abundances from the bright
GIRAFFE spectra of cluster members (membership is described above in \S 3), 
 using the same techniques described above.
The derived abundances from the GIRAFFE spectra are listed in Tables \ref{Gabund1}
and \ref{Gabund2}, together with their uncertainties.

The Na abundance was obtained from the spectral lines
at 5662-8 and 6154-60 \AA.
These features are well known to be affected by NLTE effects.
For this reason we applied an NLTE correction from \citet{Gr99} to the output
LTE abundances (see Table \ref{Uabund1}).

\begin{table}
 \centering
\vskip-0.2in
 \begin{minipage}{140mm}
  \caption{Measured Abundances for UVES stars.\label{Uabund1}}
  \begin{tabular}{rllllrll}
  \hline
ID & El &(X) & EL$_{Meas}$  & EL$_{\odot}$  & \# & [El/X] \\
\hline
158 & Fe{\hspace{0.07cm}\scriptsize\rm I}      &(H)  &  7.42$\pm$0.04  & 7.48      &  31 & $-$0.06$\pm$0.04 \\
158 & Fe{\hspace{0.07cm}\scriptsize\rm II}     &(H)  &  7.45$\pm$0.14  & 7.51      &   3 & $-$0.06$\pm$0.14 \\
158 & Na{\hspace{0.07cm}\scriptsize\rm I}(568) &(Fe) &  6.51$\pm$0.07  & 6.31      &   2 & $+$0.26$\pm$0.08$^L$\\[-0.2ex]
    &                                          &     &                 &           &     & $+$0.31$\pm$0.08$^N$\\
158 & Na{\hspace{0.07cm}\scriptsize\rm I}(616) &(Fe) &  6.17           & 6.31      &   1 & $-$0.08$^L$\\[-0.2ex]
    &                                          &     &                 &           &     & $+$0.07$^N$ \\
158 & Mg{\hspace{0.07cm}\scriptsize\rm I}      &(Fe) &  7.25           & 7.53      &   1 & $-$0.22 \\
158 & Si{\hspace{0.07cm}\scriptsize\rm I}      &(Fe) &  7.72$\pm$0.10  & 7.61      &   2 & $+$0.17$\pm$0.11  \\
158 & Ca{\hspace{0.07cm}\scriptsize\rm II}     &(Fe) &  6.34$\pm$0.07  & 6.37      &   5 & $+$0.03$\pm$0.08 \\
158 & Ti{\hspace{0.07cm}\scriptsize\rm I}      &(Fe) &  5.07$\pm$0.07  & 4.93      &   7 & $+$0.20$\pm$0.08 \\
158 & Ti{\hspace{0.07cm}\scriptsize\rm II}     &(Fe) &  4.93$\pm$0.10  & 4.96      &   2 & $+$0.03$\pm$0.11 \\
158 & Cr{\hspace{0.07cm}\scriptsize\rm I}      &(Fe) &  5.60$\pm$0.13  & 5.65      &   5 & $+$0.01$\pm$0.14 \\
158 & Cr{\hspace{0.07cm}\scriptsize\rm II}     &(Fe) &  5.64           & 5.72      &   1 & $-$0.02 \\
158 & Ni{\hspace{0.07cm}\scriptsize\rm I}      &(Fe) &  6.21$\pm$0.08  & 6.26      &   7 & $+$0.01$\pm$0.09 \\
158 & Ba{\hspace{0.07cm}\scriptsize\rm I}      &(Fe) &  1.98$\pm$0.08  & 2.45      &   2 & $-$0.41$\pm$0.09 \\
158 & $\alpha$                                 &(Fe) &                 &           &     & $+$0.04$\pm$0.08 \\[-1ex]\hline
162 & Fe{\hspace{0.07cm}\scriptsize\rm I}      &(H)  &  7.44$\pm$0.04  & 7.48      &  35 & $-$0.04$\pm$0.04 \\
162 & Fe{\hspace{0.07cm}\scriptsize\rm II}     &(H)  &  7.47$\pm$0.10  & 7.51      &   4 & $-$0.04$\pm$0.10 \\
162 & Na{\hspace{0.07cm}\scriptsize\rm I}(568) &(Fe) &  6.87$\pm$0.09  & 6.31      &   2 & $+$0.60$\pm$0.10$^L$\\[-0.2ex]
    &                                          &     &                 &           &     & $+$0.65$\pm$0.10$^N$\\
162 & Na{\hspace{0.07cm}\scriptsize\rm I}(616) &(Fe) &  6.35           & 6.31      &   1 & $+$0.08$^L$\\[-0.2ex]
    &                                          &     &                 &           &     & $+$0.23$^N$\\
162 & Mg{\hspace{0.07cm}\scriptsize\rm I}      &(Fe) &  7.55           & 7.53      &   1 & $+$0.06 \\
162 & Si{\hspace{0.07cm}\scriptsize\rm I}      &(Fe) &  7.72$\pm$0.08  & 7.61      &   2 & $+$0.15$\pm$0.09 \\
162 & Ca{\hspace{0.07cm}\scriptsize\rm II}     &(Fe) &  6.26$\pm$0.09  & 6.37      &   5 & $-$0.07$\pm$0.10 \\
162 & Ti{\hspace{0.07cm}\scriptsize\rm I}      &(Fe) &  4.99$\pm$0.09  & 4.93      &   8 & $+$0.10$\pm$0.10  \\
162 & Ti{\hspace{0.07cm}\scriptsize\rm II}     &(Fe) &  5.15$\pm$0.16  & 4.96      &   2 & $+$0.23$\pm$0.16 \\
162 & Cr{\hspace{0.07cm}\scriptsize\rm I}      &(Fe) &  5.80$\pm$0.13  & 5.65      &   3 & $+$0.19$\pm$0.14 \\
162 & Cr{\hspace{0.07cm}\scriptsize\rm II}     &(Fe) &  5.78$\pm$0.06  & 5.72      &   2 & $+$0.10$\pm$0.07 \\
162 & Ni{\hspace{0.07cm}\scriptsize\rm I}      &(Fe) &  6.29$\pm$0.07  & 6.26      &   7 & $+$0.07$\pm$0.08 \\
162 & Ba{\hspace{0.07cm}\scriptsize\rm I}      &(Fe) &  2.10$\pm$0.10  & 2.45      &   2 & $-$0.31$\pm$0.11 \\
162 & $\alpha$                                 &(Fe) &                 &           &     & $+$0.09$\pm$0.05 \\[-1ex]\hline
164 & Fe{\hspace{0.07cm}\scriptsize\rm I}      &(H)  &  7.39$\pm$0.03  & 7.48      &  49 & $-$0.09$\pm$0.03 \\
164 & Fe{\hspace{0.07cm}\scriptsize\rm II}     &(H)  &  7.42$\pm$0.10  & 7.51      &   3 & $-$0.09$\pm$0.10 \\
164 & Na{\hspace{0.07cm}\scriptsize\rm I}(568) &(Fe) &  6.36$\pm$0.07  & 6.31      &   2 & $+$0.14$\pm$0.07$^L$\\[-0.2ex]
    &                                          &     &                 &           &     & $+$0.19$\pm$0.08$^N$\\
164 & Na{\hspace{0.07cm}\scriptsize\rm I}(616) &(Fe) &  6.23           & 6.31      &   1 & $+$0.01$^L$\\[-0.2ex]
    &                                          &     &                 &           &     & $+$0.16$^N$\\
164 & Mg{\hspace{0.07cm}\scriptsize\rm I}      &(Fe) &  7.30           & 7.53      &   1 & $-$0.14  \\
164 & Si{\hspace{0.07cm}\scriptsize\rm I}      &(Fe) &  7.62$\pm$0.05  & 7.61      &   2 & $+$0.10$\pm$0.06 \\
164 & Ca{\hspace{0.07cm}\scriptsize\rm II}     &(Fe) &  6.30$\pm$0.05  & 6.37      &   9 & $+$0.02$\pm$0.06 \\
164 & Ti{\hspace{0.07cm}\scriptsize\rm I}      &(Fe) &  5.14$\pm$0.04  & 4.93      &  11 & $+$0.30$\pm$0.05 \\
164 & Ti{\hspace{0.07cm}\scriptsize\rm II}     &(Fe) &  5.23$\pm$0.08  & 4.96      &   2 & $+$0.36$\pm$0.08 \\
164 & Cr{\hspace{0.07cm}\scriptsize\rm I}      &(Fe) &  5.77$\pm$0.11  & 5.65      &   5 & $+$0.21$\pm$0.11 \\
164 & Cr{\hspace{0.07cm}\scriptsize\rm II}     &(Fe) &  5.89$\pm$0.07  & 5.72      &   2 & $+$0.26$\pm$0.08  \\
164 & Ni{\hspace{0.07cm}\scriptsize\rm I}      &(Fe) &  6.25$\pm$0.06  & 6.26      &  11 & $+$0.08$\pm$0.07 \\
164 & Ba{\hspace{0.07cm}\scriptsize\rm I}      &(Fe) &  2.52$\pm$0.01  & 2.45      &   2 & $+$0.16$\pm$0.02 \\
164 & $\alpha$                                 &(Fe) &                 &           &     & $+$0.13$\pm$0.10 \\[-1ex]\hline
165 & Fe{\hspace{0.07cm}\scriptsize\rm I}      &(H)  &  7.41$\pm$0.05  & 7.48      &  26 & $-$0.07$\pm$0.05 \\
165 & Fe{\hspace{0.07cm}\scriptsize\rm II}     &(H)  &  7.44$\pm$0.09  & 7.51      &   4 & $-$0.07$\pm$0.09 \\
165 & Na{\hspace{0.07cm}\scriptsize\rm I}(568) &(Fe) &  6.36$\pm$0.01  & 6.31      &   2 & $+$0.12$\pm$0.05$^L$\\[-0.2ex]
    &                                          &     &                 &           &     & $+$0.17$\pm$0.05$^N$\\
165 & Na{\hspace{0.07cm}\scriptsize\rm I}(616) &(Fe) &  6.43           & 6.31      &   1 & $+$0.19$^L$ \\[-0.2ex]
    &                                          &     &                 &           &     & $+$0.34$^N$\\
165 & Mg{\hspace{0.07cm}\scriptsize\rm I}      &(Fe) &  7.18           & 7.53      &   1 & $-$0.28 \\
165 & Si{\hspace{0.07cm}\scriptsize\rm I}      &(Fe) &  7.72$\pm$0.18  & 7.61      &   2 & $+$0.18$\pm$0.19 \\
165 & Ca{\hspace{0.07cm}\scriptsize\rm II}     &(Fe) &  6.30$\pm$0.07  & 6.37      &   8 & $+$0.00$\pm$0.09 \\
165 & Ti{\hspace{0.07cm}\scriptsize\rm I}      &(Fe) &  5.18$\pm$0.11  & 4.93      &   7 & $+$0.32$\pm$0.12 \\
165 & Ti{\hspace{0.07cm}\scriptsize\rm II}     &(Fe) &  5.01$\pm$0.64  & 4.96      &   2 & $+$0.12$\pm$0.64 \\
165 & Cr{\hspace{0.07cm}\scriptsize\rm I}      &(Fe) &  5.77$\pm$0.14  & 5.65      &   5 & $+$0.19$\pm$0.15 \\
165 & Cr{\hspace{0.07cm}\scriptsize\rm II}     &(Fe) &                 & 5.72      &   0 &  \\
165 & Ni{\hspace{0.07cm}\scriptsize\rm I}      &(Fe) &  6.37$\pm$0.08  & 6.26      &   4 & $+$0.18$\pm$0.09 \\
165 & Ba{\hspace{0.07cm}\scriptsize\rm I}      &(Fe) &  2.40$\pm$0.16  & 2.45      &   2 & $+$0.02$\pm$0.17 \\
165 & $\alpha$                                 &(Fe) &                 &           &     & $-$0.07$\pm$0.11 \\
\hline
\end{tabular}\\
$^N$ NLTE solution\\
$^L$ LTE solution
\end{minipage}
\end{table}

\begin{table}
 \centering
 \begin{minipage}{140mm}
  \caption{Measured Abundances for GIRAFFE stars.\label{Gabund1}}
  \begin{tabular}{rllllrll}
  \hline
ID & El &(X) & EL$_{Meas}$  & EL$_{\odot}$  & \# & [El/X] \\
\hline
 63 & Fe{\hspace{0.07cm}\scriptsize\rm I}  &(H)  & 7.32$\pm$0.05 & 7.51 &  34 &  $-$0.19$\pm$0.05 \\
 63 & Fe{\hspace{0.07cm}\scriptsize\rm II} &(H)  & 7.34          & 7.52 &   1 &  $-$0.18           \\
 63 & Ti{\hspace{0.07cm}\scriptsize\rm I}  &(Fe) & 5.09$\pm$0.15 & 5.08 &   6 &  $+$0.20$\pm$0.16 \\
 63 & Ti{\hspace{0.07cm}\scriptsize\rm II} &(Fe) & 5.08$\pm$0.15 & 5.08 &   3 &  $+$0.19$\pm$0.16 \\
 63 & Cr{\hspace{0.07cm}\scriptsize\rm I}  &(Fe) & 5.17$\pm$0.26 & 5.66 &   6 &  $-$0.30$\pm$0.26 \\
 63 & Ni{\hspace{0.07cm}\scriptsize\rm I}  &(Fe) & 6.12$\pm$0.19 & 6.51 &   4 &  $-$0.20$\pm$0.20 \\[1ex]
 98 & Fe{\hspace{0.07cm}\scriptsize\rm I}  &(H)  & 7.45$\pm$0.05 & 7.51 &  27 &  $-$0.06$\pm$0.05 \\
 98 & Fe{\hspace{0.07cm}\scriptsize\rm II} &(H)  & 7.45$\pm$0.08 & 7.52 &   3 &  $-$0.07$\pm$0.08 \\
 98 & Ti{\hspace{0.07cm}\scriptsize\rm I}  &(Fe) & 5.15$\pm$0.13 & 5.08 &   6 &  $+$0.13$\pm$0.14 \\
 98 & Ti{\hspace{0.07cm}\scriptsize\rm II} &(Fe) & 5.00$\pm$0.13 & 5.08 &   3 &  $-$0.02$\pm$0.14 \\
 98 & Cr{\hspace{0.07cm}\scriptsize\rm I}  &(Fe) & 5.16$\pm$0.18 & 5.66 &   4 &  $-$0.44$\pm$0.19 \\
 98 & Ni{\hspace{0.07cm}\scriptsize\rm I}  &(Fe) & 5.89$\pm$0.17 & 6.51 &   3 &  $-$0.56$\pm$0.18 \\[1ex]
102 & Fe{\hspace{0.07cm}\scriptsize\rm I}  &(H)  & 7.25$\pm$0.06 & 7.51 &  19 &  $-$0.26$\pm$0.06 \\
102 & Fe{\hspace{0.07cm}\scriptsize\rm II} &(H)  & 7.29$\pm$0.01 & 7.52 &   2 &  $-$0.23$\pm$0.01 \\
102 & Ti{\hspace{0.07cm}\scriptsize\rm I}  &(Fe) & 5.20$\pm$0.09 & 5.08 &   5 &  $+$0.38$\pm$0.11 \\
102 & Ti{\hspace{0.07cm}\scriptsize\rm II} &(Fe) & 5.29$\pm$0.15 & 5.08 &   2 &  $+$0.47$\pm$0.16 \\
102 & Cr{\hspace{0.07cm}\scriptsize\rm I}  &(Fe) & 5.21$\pm$0.16 & 5.66 &   3 &  $-$0.19$\pm$0.17 \\
102 & Ni{\hspace{0.07cm}\scriptsize\rm I}  &(Fe) & 6.27$\pm$0.17 & 6.51 &   2 &  $+$0.02$\pm$0.18 \\[1ex]
124 & Fe{\hspace{0.07cm}\scriptsize\rm I}  &(H)  & 7.47$\pm$0.04 & 7.51 &  25 &  $-$0.04$\pm$0.04 \\
124 & Fe{\hspace{0.07cm}\scriptsize\rm II} &(H)  & 7.47$\pm$0.17 & 7.52 &   2 &  $-$0.05$\pm$0.17 \\
124 & Ti{\hspace{0.07cm}\scriptsize\rm I}  &(Fe) & 5.35$\pm$0.13 & 5.08 &   5 &  $+$0.31$\pm$0.14 \\
124 & Ti{\hspace{0.07cm}\scriptsize\rm II} &(Fe) & 5.45$\pm$0.29 & 5.08 &   3 &  $+$0.41$\pm$0.29 \\
124 & Cr{\hspace{0.07cm}\scriptsize\rm I}  &(Fe) & 5.46$\pm$0.06 & 5.66 &   3 &  $-$0.16$\pm$0.07 \\
124 & Ni{\hspace{0.07cm}\scriptsize\rm I}  &(Fe) & 6.18          & 6.51 &   1 &  $-$0.29          \\[1ex]
127 & Fe{\hspace{0.07cm}\scriptsize\rm I}  &(H)  & 7.16$\pm$0.06 & 7.51 &  27 &  $-$0.35$\pm$0.06 \\
127 & Fe{\hspace{0.07cm}\scriptsize\rm II} &(H)  & 7.17$\pm$0.02 & 7.52 &   2 &  $-$0.35$\pm$0.02 \\
127 & Ti{\hspace{0.07cm}\scriptsize\rm I}  &(Fe) & 4.97$\pm$0.13 & 5.08 &   7 &  $+$0.24$\pm$0.14 \\
127 & Ti{\hspace{0.07cm}\scriptsize\rm II} &(Fe) & 5.31$\pm$0.50 & 5.08 &   2 &  $+$0.58$\pm$0.50 \\
127 & Cr{\hspace{0.07cm}\scriptsize\rm I}  &(Fe) & 4.74$\pm$0.19 & 5.66 &   4 &  $-$0.57$\pm$0.20 \\
127 & Ni{\hspace{0.07cm}\scriptsize\rm I}  &(Fe) & 6.19$\pm$0.18 & 6.51 &   2 &  $+$0.03$\pm$0.19 \\[1ex]
135 & Fe{\hspace{0.07cm}\scriptsize\rm I}  &(H)  & 7.43$\pm$0.06 & 7.51 &  27 &  $-$0.08$\pm$0.06 \\
135 & Fe{\hspace{0.07cm}\scriptsize\rm II} &(H)  & 7.44$\pm$0.08 & 7.52 &   2 &  $-$0.08$\pm$0.85 \\
135 & Ti{\hspace{0.07cm}\scriptsize\rm I}  &(Fe) & 5.17$\pm$0.15 & 5.08 &   6 &  $+$0.17$\pm$0.16 \\
135 & Ti{\hspace{0.07cm}\scriptsize\rm II} &(Fe) & 5.22$\pm$0.11 & 5.08 &   3 &  $+$0.22$\pm$0.12 \\
135 & Cr{\hspace{0.07cm}\scriptsize\rm I}  &(Fe) & 5.00$\pm$0.25 & 5.66 &   5 &  $-$0.58$\pm$0.26 \\
135 & Ni{\hspace{0.07cm}\scriptsize\rm I}  &(Fe) & 6.28$\pm$0.10 & 6.51 &   2 &  $-$0.15$\pm$0.12 \\[1ex]
140 & Fe{\hspace{0.07cm}\scriptsize\rm I}  &(H)  & 7.50$\pm$0.07 & 7.51 &  21 &  $-$0.01$\pm$0.07 \\
140 & Fe{\hspace{0.07cm}\scriptsize\rm II} &(H)  & 7.50$\pm$0.16 & 7.52 &   2 &  $-$0.02$\pm$0.16 \\
140 & Ti{\hspace{0.07cm}\scriptsize\rm I}  &(Fe) & 5.08$\pm$0.06 & 5.08 &   5 &  $+$0.01$\pm$0.09 \\
140 & Ti{\hspace{0.07cm}\scriptsize\rm II} &(Fe) & 4.98$\pm$0.49 & 5.08 &   2 &  $-$0.09$\pm$0.49 \\
140 & Cr{\hspace{0.07cm}\scriptsize\rm I}  &(Fe) & 5.07$\pm$0.25 & 5.66 &   4 &  $-$0.58$\pm$0.26 \\
140 & Ni{\hspace{0.07cm}\scriptsize\rm I}  &(Fe) & 6.45$\pm$0.11 & 6.51 &   2 &  $-$0.05$\pm$0.13 \\[1ex]
177 & Fe{\hspace{0.07cm}\scriptsize\rm I}  &(H)  & 7.08$\pm$0.05 & 7.51 &  20 &  $-$0.43$\pm$0.05 \\
177 & Fe{\hspace{0.07cm}\scriptsize\rm II} &(H)  & 7.09$\pm$0.12 & 7.52 &   3 &  $-$0.43$\pm$0.12 \\
177 & Ti{\hspace{0.07cm}\scriptsize\rm I}  &(Fe) & 5.00$\pm$0.18 & 5.08 &   4 &  $+$0.35$\pm$0.19 \\
177 & Ti{\hspace{0.07cm}\scriptsize\rm II} &(Fe) & 5.19$\pm$0.11 & 5.08 &   2 &  $+$0.54$\pm$0.12 \\
177 & Cr{\hspace{0.07cm}\scriptsize\rm I}  &(Fe) & 5.05$\pm$0.20 & 5.66 &   5 &  $-$0.18$\pm$0.21 \\
177 & Ni{\hspace{0.07cm}\scriptsize\rm I}  &(Fe) & 5.79$\pm$0.28 & 6.51 &   2 &  $-$0.29$\pm$0.28 \\[1ex]
179 & Fe{\hspace{0.07cm}\scriptsize\rm I}  &(H)  & 7.46$\pm$0.04 & 7.51 &  22 &  $-$0.05$\pm$0.04 \\
179 & Fe{\hspace{0.07cm}\scriptsize\rm II} &(H)  & 7.46$\pm$0.25 & 7.52 &   2 &  $-$0.06$\pm$0.25 \\
179 & Ti{\hspace{0.07cm}\scriptsize\rm I}  &(Fe) & 5.16$\pm$0.19 & 5.08 &   5 &  $+$0.23$\pm$0.19 \\
179 & Ti{\hspace{0.07cm}\scriptsize\rm II} &(Fe) & 5.16$\pm$0.36 & 5.08 &   2 &  $+$0.13$\pm$0.36 \\
179 & Cr{\hspace{0.07cm}\scriptsize\rm I}  &(Fe) & 4.59          & 5.66 &   1 &  $-$1.02          \\
179 & Ni{\hspace{0.07cm}\scriptsize\rm I}  &(Fe) &               & 6.51 &   0 &                   \\[1ex]
182 & Fe{\hspace{0.07cm}\scriptsize\rm I}  &(H)  & 7.47$\pm$0.07 & 7.51 &  20 &  $-$0.04$\pm$0.07 \\
182 & Fe{\hspace{0.07cm}\scriptsize\rm II} &(H)  & 7.48$\pm$0.39 & 7.52 &   2 &  $-$0.04$\pm$0.39 \\
182 & Ti{\hspace{0.07cm}\scriptsize\rm I}  &(Fe) & 5.25$\pm$0.17 & 5.08 &   5 &  $+$0.21$\pm$0.18 \\
182 & Ti{\hspace{0.07cm}\scriptsize\rm II} &(Fe) & 4.94$\pm$0.12 & 5.08 &   3 &  $-$0.10$\pm$0.14 \\
182 & Cr{\hspace{0.07cm}\scriptsize\rm I}  &(Fe) & 5.27$\pm$0.25 & 5.66 &   4 &  $-$0.35$\pm$0.26 \\
182 & Ni{\hspace{0.07cm}\scriptsize\rm I}  &(Fe) & 6.39          & 6.51 &   1 &  $-$0.08          \\
\hline
\end{tabular}
\end{minipage}
\end{table}
\begin{table}
 \centering
 \begin{minipage}{140mm}
  \caption{Measured Abundances for GIRAFFE stars (Cont).\label{Gabund2}}
  \begin{tabular}{rllllrll}
  \hline
ID & El &(X) & EL$_{Meas}$  & EL$_{\odot}$  & \# & [El/X] \\
\hline
191 & Fe{\hspace{0.07cm}\scriptsize\rm I}  &(H)  & 7.26$\pm$0.06 & 7.51 &  19 &  $-$0.25$\pm$0.06 \\
191 & Fe{\hspace{0.07cm}\scriptsize\rm II} &(H)  & 7.23$\pm$0.48 & 7.52 &   3 &  $-$0.29$\pm$0.48 \\
191 & Ti{\hspace{0.07cm}\scriptsize\rm I}  &(Fe) & 5.21$\pm$0.18 & 5.08 &   4 &  $+$0.38$\pm$0.19 \\
191 & Ti{\hspace{0.07cm}\scriptsize\rm II} &(Fe) & 4.99$\pm$0.26 & 5.08 &   2 &  $+$0.16$\pm$0.27 \\
191 & Cr{\hspace{0.07cm}\scriptsize\rm I}  &(Fe) & 5.29$\pm$0.07 & 5.66 &   3 &  $-$0.12$\pm$0.09 \\
191 & Ni{\hspace{0.07cm}\scriptsize\rm I}  &(Fe) & 6.35$\pm$0.11 & 6.51 &   2 &  $+$0.09$\pm$0.12 \\[1ex]
196 & Fe{\hspace{0.07cm}\scriptsize\rm I}  &(H)  & 7.23$\pm$0.06 & 7.51 &  23 &  $-$0.28$\pm$0.06 \\
196 & Fe{\hspace{0.07cm}\scriptsize\rm II} &(H)  & 7.26$\pm$0.08 & 7.52 &   3 &  $-$0.26$\pm$0.08 \\
196 & Ti{\hspace{0.07cm}\scriptsize\rm I}  &(Fe) & 5.30$\pm$0.17 & 5.08 &   3 &  $+$0.50$\pm$0.18 \\
196 & Ti{\hspace{0.07cm}\scriptsize\rm II} &(Fe) & 4.93$\pm$0.16 & 5.08 &   3 &  $+$0.13$\pm$0.17 \\
196 & Cr{\hspace{0.07cm}\scriptsize\rm I}  &(Fe) & 5.03$\pm$0.25 & 5.66 &   4 &  $-$0.35$\pm$0.26 \\
196 & Ni{\hspace{0.07cm}\scriptsize\rm I}  &(Fe) & 6.31$\pm$0.36 & 6.51 &   4 &  $+$0.08$\pm$0.36 \\[1ex]
199 & Fe{\hspace{0.07cm}\scriptsize\rm I}  &(H)  & 7.30$\pm$0.04 & 7.51 &  20 &  $-$0.21$\pm$0.04 \\
199 & Fe{\hspace{0.07cm}\scriptsize\rm II} &(H)  & 7.31$\pm$0.42 & 7.52 &   3 &  $-$0.21$\pm$0.42 \\
199 & Ti{\hspace{0.07cm}\scriptsize\rm I}  &(Fe) & 4.99$\pm$0.15 & 5.08 &   5 &  $+$0.12$\pm$0.15 \\
199 & Ti{\hspace{0.07cm}\scriptsize\rm II} &(Fe) & 5.23$\pm$0.06 & 5.08 &   2 &  $+$0.36$\pm$0.07 \\
199 & Cr{\hspace{0.07cm}\scriptsize\rm I}  &(Fe) & 5.33$\pm$0.15 & 5.66 &   4 &  $-$0.12$\pm$0.15 \\
199 & Ni{\hspace{0.07cm}\scriptsize\rm I}  &(Fe) & 6.50          & 6.51 &   1 &  $+$0.20          \\[1ex]
231 & Fe{\hspace{0.07cm}\scriptsize\rm I}  &(H)  & 7.42$\pm$0.05 & 7.51 &  23 &  $-$0.09$\pm$0.05 \\
231 & Fe{\hspace{0.07cm}\scriptsize\rm II} &(H)  & 7.43$\pm$0.11 & 7.52 &   2 &  $-$0.09$\pm$0.11 \\
231 & Ti{\hspace{0.07cm}\scriptsize\rm I}  &(Fe) & 5.46$\pm$0.12 & 5.08 &   5 &  $+$0.47$\pm$0.13 \\
231 & Ti{\hspace{0.07cm}\scriptsize\rm II} &(Fe) & 5.11$\pm$0.10 & 5.08 &   2 &  $+$0.12$\pm$0.11 \\
231 & Cr{\hspace{0.07cm}\scriptsize\rm I}  &(Fe) & 5.32$\pm$0.11 & 5.66 &   5 &  $-$0.25$\pm$0.12 \\
231 & Ni{\hspace{0.07cm}\scriptsize\rm I}  &(Fe) & 6.35$\pm$0.10 & 6.51 &   3 &  $-$0.07$\pm$0.11 \\
\hline
\end{tabular}
\end{minipage}
\end{table}

\begin{figure}
\includegraphics[width=87mm]{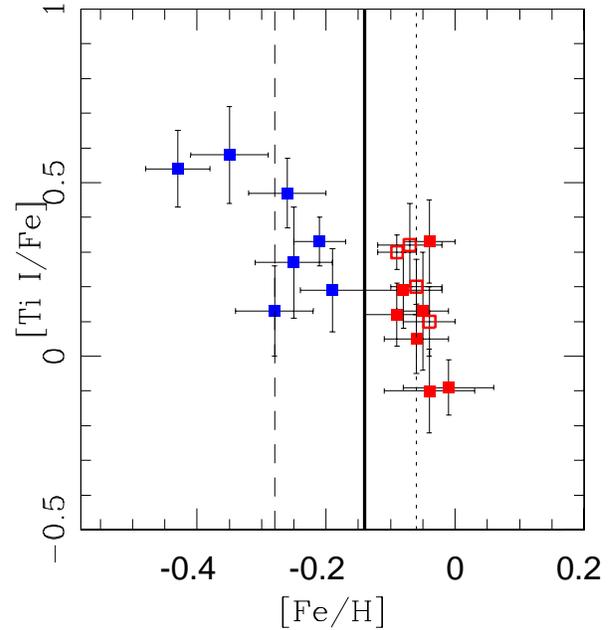}
\caption{Plot of the derived [Fe/H] vs. [Ti{\hspace{0.1cm}\scriptsize\rm I}/Fe] for the ``members'' from our analysis.
Red points denote  ``metal-rich'' and blue points are ``metal-poor'', same as Fig 1b.\label{fe_afe}.
The dotted line denotes the ``mean'' [Fe/H] value for the MR sample ($-$0.06), the
dashed line for the MP sample ($-$0.28), and the solid line for the combined sample ($-$0.14)
for the cluster if all stars are members.}
\end{figure}

\begin{figure}
\includegraphics[width=85mm]{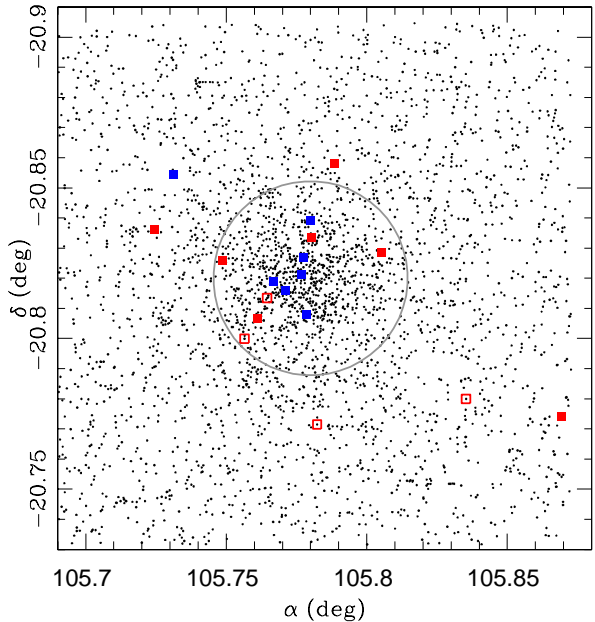}
\caption{Spatial distribution of To2 stars from \citet{phelps94} photometry.
Marked stars are ``members'' from our analysis,
with red points denoting ``metal-rich'' and blue points ``metal-poor.''  Open boxes denote the UVES stars\label{XYplot}.
The gray circle denotes a typical open clusters tidal radius of 10 pc \citep{Piskunov08} scaled to the distance of To2.  
However as To2 is in the outer disc and older, and therefore presumibly more massive than the typical open cluster, 
the shown tidal radius should be taken as a lower limit to the extent of the cluster.}
\end{figure}

\begin{figure}
\includegraphics[width=85mm]{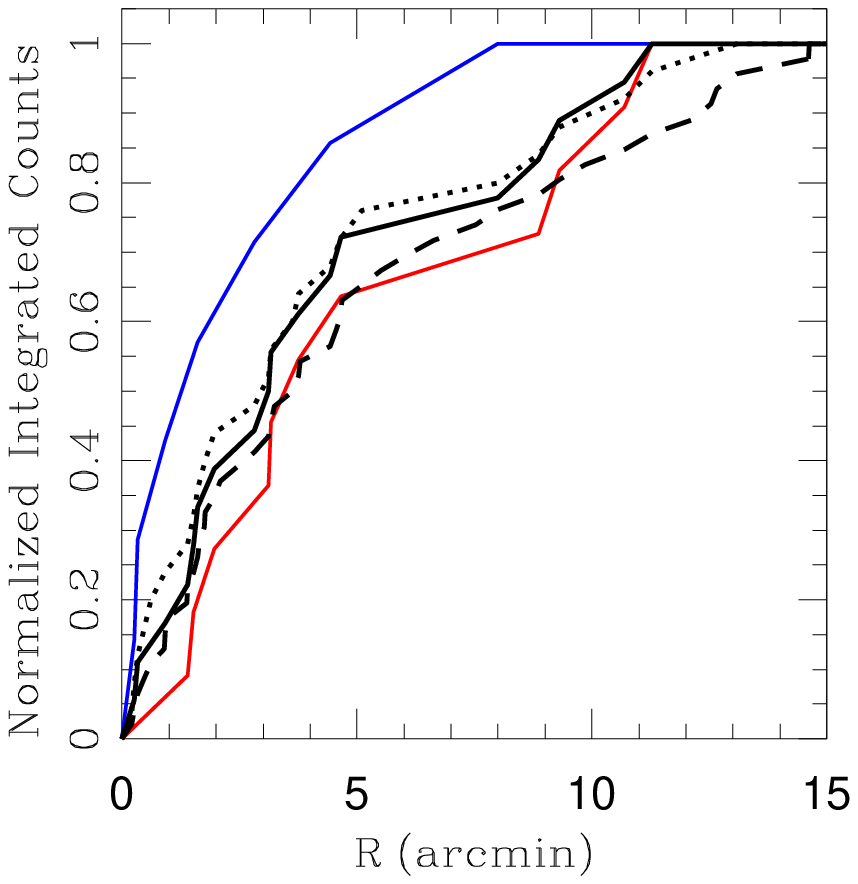}
\caption{Integrated counts of stars with RVs with 15 arcmin of the cluster centre.  
The dotted line represents all stars with RVs 
(Brown et al.\ 1996; Friel et al.\ 2000; ours), 
while the dashed line is all member stars.  The black solid line represents all of the members
found in our study \citep[excluding the stars of][]{brown96,friel02}.  
Finally, the red line show the distribution of our ``metal-rich'' members and the
blue line our ``metal-poor'' members. \label{INTGplot}}
\end{figure}

\section{Chemical Abundances and Cluster Kinematics Revisited}\label{revise}

Our abundance analysis of To2 finds a surprising spread in [Fe/H] for stars selected 
as members based on their RVs.
In Figure \ref{fe_afe}, we show the distribution of [Ti{\hspace{0.1cm}\scriptsize\rm I}/Fe] versus [Fe/H] abundance
ratios, which shows a strong overall correlation but also an apparent divide at [Fe/H] $= -0.15$ dex.
Even if the division into two populations is not real and the metallicity variation is more continuous, dividing the
sample into two parts still provides a handy way to explore properties as a function of metallicity.
To investigate this distribution further, we select two sub-samples by using the
apparent divide at [Fe/H] $= -0.15$ dex and denote these as
the ``metal-rich'' group (MR; [Fe/H] $> -0.15$ dex; red points) and the ``metal-poor''
group (MP; [Fe/H] $\le -0.15$ dex; blue points).

\begin{figure*}
\includegraphics[width=178mm]{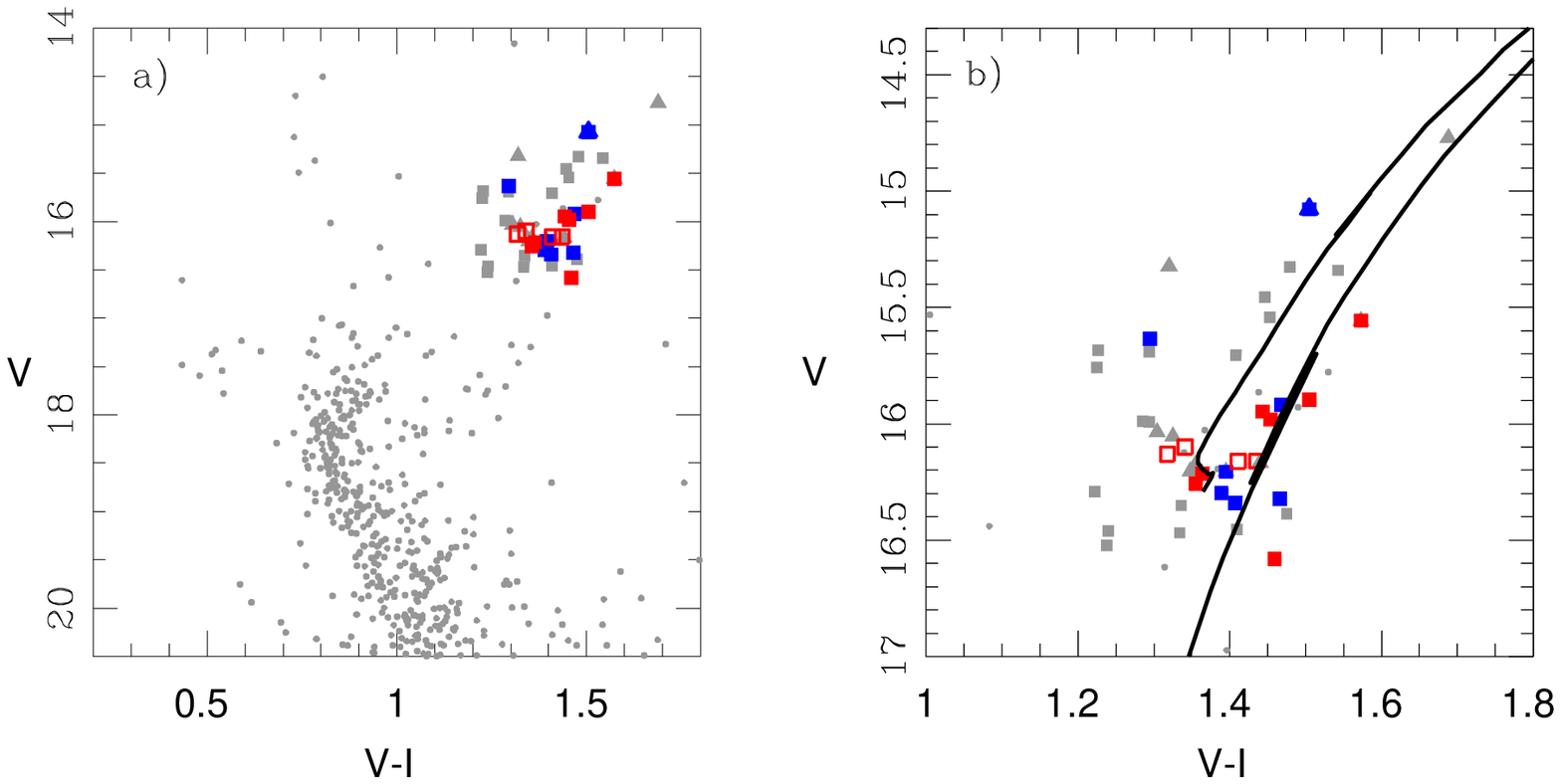}
\caption{(a) Colour-magnitude diagram for stars within a 2' radius of To2 from \citet{phelps94},
plus all stars with current and previous RV determinations.  Squares denote
observations with VLT/FLAMES+GIRAFFE, open squares with VLT/FLAMES+UVES, and triangles 
are previous observations \citep{brown96,friel02}.  
Red and blue symbols are members based on their RV.  All symbols same as Figure 1.
(b) RGB and red clump from (a) with a Padova isochrone of $Z = 0.0164$, age = 2.0 Gyr, $d = 8.4$ kpc,
and $E(B-V$) = 0.25 overplotted. \label{cmd}}.
\end{figure*}

The spatial distribution of the two sub-samples is shown in Figure \ref{XYplot} as well as in Figure \ref{rv1}b.
We find that the MP group is more centrally concentrated, as shown by the integrated counts in Figure \ref{INTGplot}, 
though both groups could reasonably belong to the cluster.  Even if we account for the selection effect 
of how stars were targeted for spectroscopy (the dotted line in Fig.\ 5)
the MP stars still show a significant central concentration.
The summary of kinematics and abundances for these sub-groups are listed in Table \ref{compare},
showing [Fe/H], [Ti{\hspace{0.1cm}\scriptsize\rm I}/Fe], [Cr/Fe], [Ni/Fe],
as well as the mean velocity ($\langle V_{r}\rangle$) and velocity dispersion ($\sigma_{int}$).
As may be seen from Table \ref{compare}, the
kinematical properties of the two samples are basically indistinguishable,
so that kinematics do little to help understand the origin of the
metallicity spread/differences among the RV-members.
Figure \ref{cmd} shows the distribution of MP and MR stars in the colour-magnitude
diagram (CMD) for To2, using the photometry from \citet{phelps94} and demonstrates that
both populations also lie in areas of the CMD expected for member stars.

\subsection{Stellar abundance reliability}

Given that an abundance spread in an open cluster is an unprecedented find, one of the first points 
to explore is whether our derived metallicities, and the implied spread in
chemistry for stars with the RV of To2, could be artificially induced by errors in our
analysis.  Two possible sources of errors 
can mimic a spread in metallicity:
\begin{enumerate}
\item Differential reddening could alter colours an unknown amount, thereby affecting
   the derived temperature, producing a spread in metallicity.  
   However, for differential reddening to explain the observed abundance spread across both 
   samples (0.4 dex) would require reddening differences of 
   $>0.2$ $E(B-V)$, which is not possible given the observed CMD and our spectral analysis.
   Moreover, Figure 4 shows that the MP and MR stars are mixed in their spatial distribution.
   In any case for our study the temperatures derived from photometry were used only as
   first guess in an iterative procedure that allowed us to find the
   atmospheric parameters directly from spectral data, and this spectral analysis reveals a reddening 
   spread of only $\delta E(B-V) \sim 0.04$ (\S \ref{atmparam}).
   As a result, we can 
   exclude differential reddening as a possible source for the observed
   metallicity spread in To2.

\item Random errors in temperature can introduce random errors in the
   derived metallicity.  However, such errors are difficult to estimate for our To2 data
   both because of the relatively small numbers of stars in our sample, and because
   of the suspected intrinsic metallicity spread of our stars.  
   The determination of the random error from $T_{eff}$ and that of the abundance determinations are coupled, 
   so that if the metallicity spread is true it would lead to an overestimation
   of the random error due to the $T_{eff}$ uncertainty.
   To get around this problem, we therefore compare our derived $T_{eff}$ uncertainty to another analyzed cluster 
   without an abundance spread to evaluate the reliability of our abundance determinations. 
   We estimated the whole error (the error due to the equivalent width
   measurement and given by MOOG + the error due to the $T_{eff}$
   uncertainty) referring to a study on the globular cluster M4 (Marino et al., in preparation).
   Some of the authors of the present study of To2 also
   derived chemical
   abundances for a large statistic sample of stars in M4
   using the same method adopted here on spectra having the same wavelength
   range and comparable $S/N$.
   Assuming a homogenous composition for the stars in M4 (an assumption
   usually
   true for most of the elements in a globular cluster with the exception
   of the light ones) a dispersion ($\sigma$) was found in the iron content of
   0.06 dex for GIRAFFE data and of 0.05 dex for UVES ones.
   These observed dispersions can be considered as a good estimate of the
   whole random error to be used in our study of To2.
   In Table \ref{errorm4} we report the dispersion for all the elements considered in this
   study as derived from the M4 data.
   Given that our measured uncertainties on the abundance determination are consistent between the M4 and To2 analysis,
   we find that errors in the determination of $T_{eff}$ cannot be the source of the observed abundance spread.
\end{enumerate}

We find that random observational errors cannot explain the mean difference of 0.22
dex (see Table \ref{compare}) in metallicity observed in To2 between the metal rich and
the metal poor group.
Additionally, since systematic errors introduced by the method affect all the
results in the same way, these also cannot explain the observed
abundance spread.
At this point we conclude that the observed spread in
metallicity in our sample is real.

\begin{table}
 \centering
 \begin{minipage}{80mm}
  \caption{Measured Abundance Ratios for UVES stars.\label{errorm4}}
  \begin{tabular}{@{}lcc@{}}
  \hline
                &  $\sigma$ from M4 & $\sigma$ from M4\\
   Element      &  (GIRAFFE) & (UVES)\\
 \hline
 error (FeI)     &    0.06   &  0.05 \\
 error (FeII)    &    0.06   &  0.05 \\
 error (TiI)     &    0.06   &  0.04 \\
 error (TiII)    &    0.10   &  0.06 \\
 error (CrI)     &    0.10   &  0.05 \\
 error (CrII)    &           &  0.08 \\
 error (Ni)      &    0.07   &  0.03 \\
 error (SiI)     &           &  0.05 \\
 error (CaI)     &           &  0.03 \\
 error (BaII)    &           &  0.03 \\

\hline
\end{tabular}
\end{minipage}
\end{table}

\subsection{Comparison to previous findings}

\citet{brown96} conducted a high resolution (R $\sim$ 34,000) 
study of To2 with the CTIO 4-m telescope 
and found [Fe/H] = $-$0.4 $\pm$ 0.25 for $E(B-V)=0.4$ or 
[Fe/H] = $-$0.5 $\pm$ 0.25 for $E(B-V)=0.3$ based on three stars, 
with [Fe/H] = $-$0.2, $-$0.4, and $-$0.57.  
The \citet{brown96} abundance analysis showed that To2   
has a reddening of
$E(B-V$) = 0.3--0.4, 
which is consistent with our spectroscopically measured reddening.  Unfortunately 
we do not have stars in common with the \citet{brown96} sample for a direct comparison
of our metallicity derivations, but the metallicities of their three stars are within the range of
our derived [Fe/H], accounting for the errors, for To2 members.
However, the more metal-rich of the three Brown et al. stars is ruled out as a cluster member based on
our radial velocity criterion.

Another study of cluster metallicities including To2 was conducted by \citet{friel02} 
using the  CTIO 4-m/ARGUS, which yielded a much 
lower resolution ($R \sim 1300$) than either our study or that of \citet{brown96}, and 
with metallicities determined from spectral indices.  
\citet{friel02} found [Fe/H] = $-$0.44 $\pm$ 0.09 for To2  
from a sample of 12 member stars, and with the individual measurements ranging from 
$-$0.28 to $-$0.65. 
\citet{friel02} found nearly identical metallicities for the two stars in common with \citet{brown96}. 
We observed three other 
\citet{friel02} stars with the GIRAFFE spectrograph but find 
poor agreement with the previous low-resolution results in 2 of the 3 stars, 
as shown in Table \ref{compareFriel}, where ID is from \citet{phelps94} and ID2 is from \citet{kubiak92}. 
It should be pointed out that sometimes high-resolution studies to find slightly 
higher metallicities when compared with low-resolution 
studies based on spectral indices (e.g., in the high resolution study of Berkeley 17, Friel et al. 2005 find 
[Fe/H] = $-0.10 \pm 0.09$ as compared to [Fe/H] = $-0.33 \pm 0.13$ from the 
low-resolution spectroscopy of Friel et al. 2002).

\begin{table*}
 \centering
 \begin{minipage}{175mm}
  \caption{GIRAFFE Spectra Radial Velocity and Abundance ``Sub-Populations''}
  \label{compare}
  \begin{tabular}{lrcccccc}
  \hline
  Sample &  \# Stars &[Fe/H] & [Ti{\hspace{0.1cm}\scriptsize\rm I}/Fe] & [Cr/Fe] & [Ni/Fe] & $\langle V_{r}\rangle$ & $\sigma_{int}$ \\
  &   &  & & & &(km s$^{-1}$) & (km s$^{-1}$) \\
\hline
Metal-rich (UVES only)   &   4 & $-$0.07$\pm$0.01 & $+$0.23$\pm$0.06 & $+$0.15$\pm$0.05 & $+$0.08$\pm$0.04 & 119.2$\pm$1.0 & 2.1$\pm$0.7 \\
Metal-rich (GIRAFFE only)&   7 & $-$0.05$\pm$0.02 & $+$0.14$\pm$0.05 & $-$0.48$\pm$0.12 & $-$0.20$\pm$0.09 & 121.7$\pm$0.5 & 1.2$\pm$0.3 \\
Metal-rich (UVES+GIRAFFE)&  11 & $-$0.06$\pm$0.01 & $+$0.23$\pm$0.03 & $-$0.25$\pm$0.07 & $-$0.05$\pm$0.05 & 120.8$\pm$0.6 & 2.0$\pm$0.4 \\
Metal-poor (GIRAFFE)     &   7 & $-$0.28$\pm$0.03 & $+$0.32$\pm$0.06 & $-$0.26$\pm$0.06 & $-$0.01$\pm$0.07 & 121.5$\pm$0.5 & 1.4$\pm$0.4 \\
All RV Members           &  18 & $-$0.14$\pm$0.03 & $+$0.24$\pm$0.03 &                  &                  & 121.0$\pm$0.4 & 1.8$\pm$0.3 \\
\hline
\end{tabular}
\end{minipage}
\end{table*}

\begin{table*}
 \centering
 \begin{minipage}{175mm}
  \caption{Comparison of \citet{friel02} and GIRAFFE Data}
  \label{compareFriel}
  \begin{tabular}{lccccccc}
  \hline
  ID & ID2 & $V_{r,GIRAFFE}$ &  $V_{r,Friel}$&  $\Delta$ $V_r$  &    [Fe/H]$_{GIRAFFE}$ & [Fe/H]$_{Friel}$ & $\Delta$[Fe/H] \\
     &     &(km s$^{-1}$)    & (km s$^{-1}$) &   (km s$^{-1}$) &                        &                  &          \\
\hline
63  &  20 &  120.96 $\pm$  0.16 &  123 $\pm$  10 &  2 & $-$0.19$\pm$0.05 & $-$0.40$\pm$0.10 &  0.21 \\
98  &  28 &  121.71 $\pm$  0.18 &  132 $\pm$  10 & 10 & $-$0.06$\pm$0.05 & $-$0.48$\pm$0.22 &  0.42 \\
177 &  61 &  118.74 $\pm$  0.16 &  139 $\pm$  10 & 20 & $-$0.43$\pm$0.05 & $-$0.39$\pm$0.31 &  0.04 \\
\hline
\end{tabular}
\end{minipage}
\end{table*}

\subsection{Revised reddening and distance}

\begin{figure*}
\includegraphics[width=179mm]{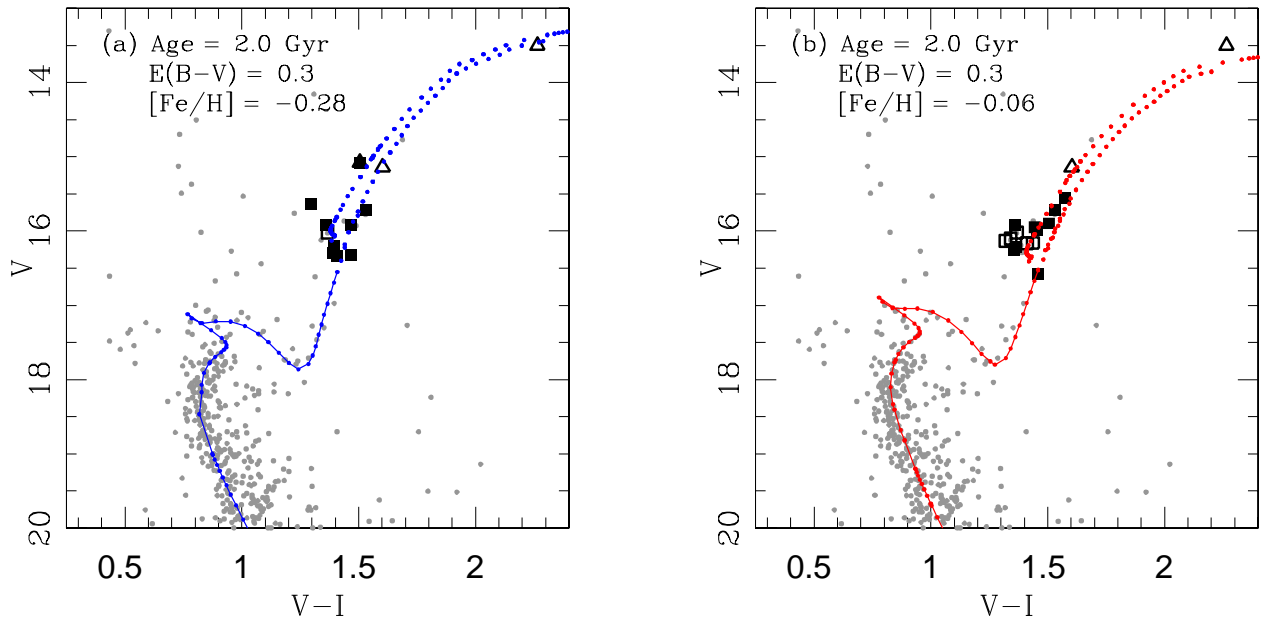}
 \caption{CMDs for To2 showing best fit Padova isochrones.
(a) Padova $Z = 0.00955$ ([Fe/H] = $-$0.28) isochrone fit is shown with
MP members and other RV members without metallicity determinations.
(b) Padova $Z = 0.01665$ ([Fe/H] = $-$0.06) isochrone fit is shown with
MR members and other RV members without metallicity determinations.\label{cmd2}
}
\end{figure*}

From the stars that have been selected as members, we have determined the 
cluster mean metal abundance for the MP and MR sub-samples
(see Table \ref{compare}).  We have generated isochrones for the exact metallicity
of each cluster sample, transforming the mean [Fe/H] into $Z$, using
Padova models and following \citet{carraro99}.  
The corresponding by-eye fit isochrone is then superposed on the
CMD, with stars of the other abundance sample removed, 
and fitted to the data distribution (Figure \ref{cmd}).
In Figure \ref{cmd2},
we match Padova isochrones to the ($V$, $V-I$) CMDs from \citet{phelps94}.
We show these matches with stars from the inner 2 arcmin of the clusters plus
RV members selected from the MP (Figure \ref{cmd2}a) and MR (Figure \ref{cmd2}b) samples.
Estimates of the basic parameters (age, distance and reddening)
are derived from comparison to the Padova isochrones. Our best fit for both samples
is: age = 2.0 Gyr, $E(B-V) = 0.3$, and $(m-M)_0 = 14.5$ or $d = 7.9$ kpc and $R_{GC} = 14.2$ kpc.
Additionally, the spectroscopy yields a reddening in the range 0.3--0.35 which is
consistent with isochrone fitting to both samples; however, we find
the value of  $E(B-V$) = 0.40 found by \citet{kubiak92}
cannot provide a good fit to the ($V$, $V-I$) CMDs using the \citet{phelps94} photometry.
Our spectroscopically determined reddening is also consistent with a newer near-infrared 
study \citep{kyeong00} who found $E(B-V$) = 0.24 $\pm$ 0.12.

\section{Discussion}

While this study was intended to verify or refute the connection of To2
with the Monoceros stream/Galactic anticenter stellar structure (GASS),
our work has produced more questions than answers.  The observed spread in To2
metallicity and $\alpha$-abundances was not expected and cannot be simply explained.
To date, no other open cluster is known to have a metallicity spread, and it is difficult 
to understand how a low-mass open cluster might have been able to retain gas after an 
initial starburst to self-enrich.
Next, we review our findings and possible scenarios that could explain the data

\subsection{Are both populations part of Tombaugh 2?}

First, we ask the question of whether both the MP and MR groups among the To2 RV members
are likely to be truly part of the cluster, and then investigate the likelihood that either population
is simply contamination of the RV member sample.

\subsubsection{Is the metal poor population part of Tombaugh 2?}

To2 has been proposed to be part of the putative GASS cluster
system \citep{pmf04,martin04a},  
and the MP population is similar, though slightly more metal rich
than, other studied Monoceros candidate clusters such as
Saurer 1 \citep{fp02,carraro03,carraro04}, Berkeley 29 \citep{carraro04,yong04},
and Palomar 1 \citep{rosen98,ata07}.  These clusters all have ages of approximately 4--6 Gyr,  metallicities
between $-$0.7 and $-$0.4 dex, and 
are all $\alpha$--enhanced.  Our To2 MP group is younger though similarly moderately metal poor
([Fe/H] $\sim$ $-$0.3) and also $\alpha$--enhanced.  This is at least circumstantial evidence
that the MP group might be the more likely member population.
This subgroup also has a metallicity roughly consistent with 
previous metallicity determinations for To2 by 
\citet{brown96} and \citet{friel02} --- though it is not impossible that these
other researchers could have also been hit with a spate of bad luck in terms
of misleading target selection.

On the other hand, it must also be admitted that the MP group both
exhibits more scatter in terms of metallicity properties than the MR population and
has a smaller number of members --- though this number increases when we account for 
the additional ``MP''
members found in the \citet{brown96} and \citet{friel02} samples that are not part of our sample.
Perhaps the strongest support for the 
supposition that the MP group is a ``real'' part of the cluster comes from the fact that 
the MP group is strongly centrally concentrated (Figs. 4 and 5) and that it
has a marginally smaller velocity dispersion than the MR group, as shown in Table \ref{compare}.

\subsubsection{Is the metal rich population part of Tombaugh 2?}

The MR population is indeed less centrally concentrated that the MP population, and in
fact has a distribution function not unlike our spectroscopic sampling function, as shown in 
Figure \ref{XYplot} and \ref{INTGplot}.  Given the small number statistics overall, it
may not be impossible for us to have been successful at achieving a uniform ``success'' 
fraction (i.e. in identifying RV members) over the radial range sampled, even by chance.
Indeed, the overall distribution of RV members very closely tracks the sampling function.
Though Figure \ref{XYplot} clearly shows a tighter MP distribution than MR population, it is
not inconceivable for there to exist a metallicity gradient in the system to yield a differential
radial distribution --- on the other hand, the one needed (i.e. more metal poor towards the centre) 
is quite unusual and not seen elsewhere.

On the other hand, the MR group has more members and is more tightly clumped in [Fe/H].  Given the
tight RV distribution, chemical properties, and sheer numbers it seems
difficult to believe that the MR population is not some coherent Simple Stellar Population (SSP), whether part
of the cluster or a ``contaminant.''  We show below that this population is quite
unlike that of the disc field star population (which is not an SSP). 

If the true To2 stars are those in the MR group (and excluding the MP group)
then To2 is less consistent with being part of the putative GASS cluster system, as it would 
be inconsistent with the observed metallicity distribution of Monoceros 
(Crane et al 2003; Sbordone et al 2005; de Jong et al. 2007). 
In this case, To2 would be the most distant metal-rich Milky Way disk open cluster found to date.  
It seems unlikely that a 2 Gyr old cluster with near solar [Fe/H] could have formed ``normally'' in 
the outer disc, given the observed, more depleted 
chemistry of the surrounding disc stars
\citep[e.g.,][]{yong2,yong3}; this would suggest that the cluster must have formed elsewhere
(e.g., in a dwarf galaxy) and been deposited in the outer disc.  This is not out of line
with current CDM models for the formation of Milky Way-like galaxies, which suggest
that discs form from the outside via continued merging (e.g., Abadi et al 2003).  
Stars as metal rich as our MR group are indeed found in paradigm accretion events, like the 
Sagittarius dwarf (e.g., Smecker-Hane \& McWilliam 2002, Monaco et al. 2005, Chou et al. 2007).

\subsubsection{Can either population simply be contamination?}

If we believe that only one of the two metallicity populations is truly part of To2, then
the other population must be a contaminant of some kind.  Here we address how likely
it is that either population may be part of the Galactic disc.

We have already seen (\S \ref{model}) that the
predicted contamination rate of normal disc stars into our RV membership range  
based on the Besancon model (see \S \ref{model}) is already smaller (by factor of 10$^2$) than
either our MP or MR samples --- thus, it seems very unlikely that either population
can be a statistical, ``fluke'' superposition of a large number of disc stars with just the right RV.
From Poisson statistics, the chances that we would get 7 (e.g., MP) or 11 (e.g., MR) 
stars as contaminants when the models predict 0.02 stars
is $1.1 \times 10^{-5}$ \% and $2.3 \times 10^{-11}$ \%, respectively.  
And this ignores the fact that other
studies have identified additional stars in the To2 that we would categorize
as ``MP.''

Based on the radial distribution of members, the MP population seems more likely
to be truly part of the cluster.  This could then imply that
the MR group is a contaminant population.  However, stars this metal rich 
(nearly solar metallicity) 
at $R_{gc} \ge 15$ kpc are not consistent with the measured metallicities
of red giants in the outer disc \citep{yong2} as well as outer disc Cepheid stars \citep{yong3},
which both suggest that the median disc metallicity at this Galactocentric distance should
be [Fe/H] $\sim$ -0.4.
These studies of outer disc tracers
find no stars as metal rich as our MR group, even for the younger Cepheid populations.

This then might suggest that the MP group could be a contaminating
field population, since the chemistry of this group is more in line
with that of the
\citet{yong2} and \citet{yong3} stars in the outer disc.  In turn, this implies
that the tight radial concentration of the MP stars is a statistical anomaly.
However Monte Carlo simulations show that the probability for a random disc 
population to achieve the observed level of concentration, 
taking into account our sampling function, is 0.1\%.  
It also implies that previous spectroscopic studies had similarly ``bad luck'' in 
picking targets to represent To2.

\subsection{Does Tombaugh 2 really have a metallicity spread?}

The cumulative evidence for To2 --- the nearly identical velocities and velocity dispersions of the stars
when divided into MP and MR groups, the positions of the stars in the CMD, the unlikeliness
that these stars are field contamination, etc. ---
suggests that all of the RV ``members'' in our To2 spectroscopic
sample could truly be members of the system.
If all RV ``members'' are indeed members of To2 then we are finding evidence for a {\it significant} metallicity spread
within To2.  That we also seem to find a similar bifurcation of [Ti{\hspace{0.1cm}\scriptsize\rm I}/Fe] with [Fe/H], rather than
a scatter-diagram in Figure 3 is further evidence of a possible coherent chemical enrichment history.
It is also possible, based on an apparent ``break'' in both [Fe/H] {\it and} [Ti{\hspace{0.1cm}\scriptsize\rm I}/Fe], that
the To2 chemical and age distribution suggests the possibility that there may be two populations
rather than a ``trend.''

If these populations of varying chemistry 
do belong to To2, what formation scenarios could create such a unique open cluster?

\subsubsection{Is To2 actually two ``very close'' clusters?}

What is the possibility that two clusters with the same RV lie
at nearly the same distance and have nearly the same
metallicity and could also be aligned along the same line of sight?
Though such a configuration could produce the observed results in the To2 field,
and even account for the apparent ``bimodal'' behavior of the chemical distribution, 
it seems extremely unlikely given the small number of clusters 
(less than 20) known this far from the Galactic centre.
It is the case that a superposition of clusters would increase the ``discoverability''
of this distant ``double cluster'', 
but the individual clusters would have to be very close together along the line of sight given that
a significant broadening 
of the main sequence is not seen in the available photometry.
However new higher quality photometry would be needed to fully test this 
case \citep[as was done with the 
{\it Hubble Space Telescope} in the case of NGC 2808;][]{n2808}.

There are a few cases of overlapping clusters, with the most-well studied overlapping pair being
NGC 1750 and NGC 1758, with ages of 200 $\pm$ 50 and 400 $\pm$ 100 Myr respectively \citep{gjt98b}.
For NGC 1750 and NGC 1758 proper motions were required to be able to separate 
the two clusters separated by $\sim 130$ pc along the line of sight \citep{gjt98b}, but with little 
noticeable differences seen in the overlapping CMD sequences \citep{gjt98a}. 
These clusters are the oldest pair of confirmed overlapping clusters, and, given the age and distance 
differences, \citet{gjt98b} conclude that the clusters do not constitute a binary system.
However for To2, the fact that mean velocity derived from high-resolution RVs is the same, 
and different than the Galactic disc trend, 
makes the possibility of it being part of a ``double cluster'' remote.

\subsubsection{Does To2 represent the merging of two clusters?}

Another explanation for our results in the To2 field might be that
two clusters with nearly the same age and different metallicities could have have collided and merged.  
Given that open clusters rotate with the Galactic disc, the small relative velocity difference between 
two clusters with similar orbits makes 
prolonged near encounters between close clusters possible.  
Additionally, the relatively large velocity dispersion of To2,
 $\sigma_{int} = 1.8$ km s$^{-1}$
(e.g., M67 has $\sigma_{int} = 0.8$--0.96 km s$^{-1}$; Girard et al. 1989; Frinchaboy \& Majewski 2007) 
might suggest 
that the cluster could have been dynamically heated at some point, possibly by a merger of clusters.
A simple calculation based on merging of two equal-mass eliptical galaxies \citep{bt}, 
shows that for a two-body encounter 
with two equally massed clusters (assuming each is $10^4$ M$_{\sun}$) finds that
mergers are possible if two clusters 
pass within less than 2 pc given a differential velocity of 5 km s$^{-1}$.
However as pointed out in the argument above,   
given the small number of distant clusters
the chances of such an occurrence are very small; nevertheless, 
the merging of two clusters cannot be ruled out as a possible explanation for
the observed chemical peculiarities.

\subsubsection{Has To2 had multiple episodes of star formation?}

The cumulative evidence for To2 --- the nearly identical velocities and velocity dispersions of the stars
when divided into MP and MR groups, the positions of the stars in the CMD, the unlikeliness
that these stars are field contamination, etc. ---
suggests that all of the RV ``members'' in our To2 spectroscopic
sample could truly be members of the system.
If all RV ``members'' are indeed members of To2 then we are finding evidence for a {\it significant} metallicity spread
within To2.  That we also seem to find a similar bifurcation of [Ti{\hspace{0.1cm}\scriptsize\rm I}/Fe] with [Fe/H], rather than
a scatter-diagram in Figure 3 is further evidence of a possible coherent chemical enrichment history.
It is also possible, based on an apparent ``break'' in both [Fe/H] {\it and} [Ti{\hspace{0.1cm}\scriptsize\rm I}/Fe], that
the To2 chemical and age distribution suggests the possibility that there may be two populations
rather than a ``trend.''

No other open cluster has been found to have an internal metallicity spread.
On the other hand, a small number of {\it globular} 
clusters have been found with metallicity spreads (e.g., $\omega$ Cen;  
Norris, Freeman, \& Mighell 1996; Majewski et al. 2000; 
Carraro \& Lia 2000; Frinchaboy et al. 2002; Villanova et al. 2007), 
multiple internal populations \citep[NGC 1851 and NGC 2808; Milone et al. 2007,][]{n2808}, or to represent one of a series of populations
in a more complex structure (M54 in the Sagittarius dwarf spheroidal galaxy; Sgr dSph; Siegel et al. 2007).
Interestingly, such unusual clusters typically have been 
associated with the possibility of tidal accretion events.

Could To2 be the remains of something that was once larger --- a globular cluster or dwarf galaxy?
This seems like the only way
that To2 could have had two episodes of star formation, since the mass of a typical open
cluster is generally much too small to retain and self-enrich gas after an initial
star burst.
The mass of a long-lived open cluster is typically a few $10^3$--$10^4$ M$_{\sun}$ 
\citep[e.g., M67 has a current mass $\sim 2 \times 10^3 \; {\rm M}_{\sun}$;][]{hurley05}.
A mass of at least $10^7$ M$_{\sun}$ would be needed for To2 to have retained 
enough gas to have a second epoch of star formation 
(e.g., the current mass of $\omega$ Cen is $2.8 \times 10^6$; D'Souza \& Rix 2005).
Thus, this would require To 2 either to have had a significant dark matter content, 
which is ruled out by the velocity dispersion (and in any case would
have made To 2 unique among star clusters, open or globular), or
to have been initially much larger and then stripped 
down to its current size.
Fortunately for this scenario,
the disc is a dynamically brutal environment for clusters and dwarf galaxies, which are expected to
undergo vigorous stripping.  Indeed, this is the prevailing scenario to explain the presently
visible multiple populations in the $\omega$ Cen ``globular cluster'' system.

The two apparent sub-populations in To2 do seem to show a chemical trend evoking 
self-enrichment, with the metal-poor, $\alpha$-enhanced 
population formed first, followed by formation of a more metal-rich, non-$\alpha$-enhanced population.  
The above ``evolution''
is what is expected from closed-box chemical evolution models.
Unfortunately, two other observational facts contradict the self enrichment
scenario: (1) The derived isochrone ages for the two populations are the same, within the errors of the
fitting, which are $\pm$ 0.5 Gyr.
(2) It is typical in self-enrichment scenarios for the more metal rich population, formed in the
bottom of the potential well more recently, to be more centrally concentrated.  Thus, 
if To2 was the tidal remnant of a larger stellar system one would expect that the ``younger'' metal-rich population
would be more centrally concentrated, as is found in the case of $\omega$ Cen (Pancino et al. 2000) 
and dSph galaxies (e.g., Sculptor; Tolstoy et al 2004).  However given the available data, 
we find that the metal-poor stars are more centrally concentrated,
in marked contrast to large systems undergoing tidal stripping (Tolstoy et al 2004).

\subsubsection{A ``hybrid'' solution?}

We have seen how a self-enrichment scenario seems to be contradicted by the relative
radial and age distributions of the MP and MR populations.  And we have seen that it is 
unlikely that To2 represents either dynamically or apparently merged/overlapped systems.
Yet the strongly matched velocities and velocity dispersions of the two systems suggests
a dynamical connection of the two populations, and there is at least some metallicity and
velocity mismatch of these populations with the surrounding
disc stars.  There remains one possible scenario that could account for most of the
observed properties in the To2 field, and which has a known prototype: The Sgr dSph 
stream and its cluster system.  

Any small spectroscopic survey of a Sgr star cluster, particularly one of those Sgr
clusters near the core of the dSph (e.g., Terzan 7, Terzan 8 or Arp 2),
would produce a number of the same phenomena we observe in our small spectroscopic survey of the
To2 field:  A centrally concentrated, older core of more metal-poor stars that are part of
the star cluster, seen against a more broadly distributed, typically younger and more 
metal rich population of stars, generally unbound, tidally stripped from the parent dSph
galaxy, and all with the same radial velocity.  This could be an analogous situation to what
we observe in the field of To2, with the primary difference being that we are explaining 
the properties of an open, not globular, cluster.

Thus, we hypothesize this as the leading explanation for what we are finding in the To2 
field:  We propose that the MP population represents the true cluster, as most strongly
supported by the radial concentration of these stars, and that the MR population represents
stars that are part of a distinct parent system, probably a tidally disrupted dwarf galaxy, that are 
spread more uniformly across our survey field (and beyond).  Such a scenario is consistent
with prevailing models of disc formation, which indeed suggest that discs continually grow
by accreting ``subhalos'' around the edge (e.g., Abadi et al 2003, Brook et al. 2004). 
 It also accounts for how we
can have two populations with identical RVs, but differing metallicities, and with the more
metal-rich population being more extended.  That To2 has been already hypothesized
to be a part of the GASS/Monoceros ``tidal stream'' \citep{pmf04,martin04a} provides additional support for this
proposed scenario, although the velocity dispersion of GASS/Mon is much larger than our MR population
(Crane et al. 2003, Martin et al 2006). 
Nevertheless, it is possible that other tidal streams could exist and have contributed
star clusters to the outer disc.

\subsection{Final remarks}

While we have discussed
 a number of possible explanations for our unusual findings in our survey of the To2 field, the
 only hypothesis that seems to explain most of the results in a natural way is that To2 is 
 represented by our MP population and that the MR population represents stars from a 
 parent dwarf galaxy that is contributing To2 to our Galactic disc.  All other hypotheses ---
 self-enrichment, a merger of two clusters, a superposition of two clusters --- have serious
 problems or are very improbable.  It is also yet unclear whether To2 may be part of the  proposed GASS/Mon system. 
Surely it is worth mentioning that if To2 actually has two different populations, then it is truly an unusual object. 
The one clear finding from this work is that
further work is needed on the outer Galactic cluster system, in particular To2.  A much larger spectroscopic
sample that includes both high precision velocities and detailed abundances would be 
especially useful. Moreover, the properties
of outer disc stars are needed to better constrain the existence, characteristics and 
origin of the GASS/Monoceros system and other tidally accreted systems
in the outer Galactic disc.
While RVs and abundances may not provide an explicit separation of an accreted satellite from the outer disc,
it is nonetheless essential that the kinematical and more importantly abundance trends in the outer disc
are known.  Without clear knowledge of the abundance patterns with radius,
especially at large distances and in the second and third
Galactic quadrants, one cannot explicitly use abundances to distinguish whether the GASS/Monoceros stream
is distinct from the outer Galactic disc.  The present paper is one among many efforts 
that aim to correct the
deficiency of abundance information known for the outer Milky Way disc.

\section*{Acknowledgments}

PMF is supported by
a National Science Foundation
(NSF) Astronomy and Astrophysics Postdoctoral Fellowship under award AST-0602221.
D.G. gratefully acknowledges support from the Chilean
{\sl Centro de Astrof\'\i sica} FONDAP No. 15010003.
SRM appreciates support from NSF grant
AST-0307851, NASA/JPL contract 1228235 and the David and Lucile
Packard Foundation as well as Frank Levinson through the
Peninsular Community Foundation.

\label{lastpage}
\end{document}